# VIRTUAL AND AUGMENTED REALITY-BASED ASSISTIVE INTERFACES FOR UPPER-LIMB PROSTHESIS CONTROL AND REHABILITATION

by

Yinghe Sun

A thesis submitted to Johns Hopkins University in conformity with the requirements for the degree of Master of Science in Engineering

Baltimore, Maryland
May, 2021

# Abstract

Upper-limb motor deficits present individuals with physical and psychological barriers which can significantly impede one's ability to independently continue with activities of daily living. While conventional therapeutic interventions help patients to regain functionality and mobility, many of those are often monotonous and less engaging, resulting in insufficient patient motivation and dedication to complete the regimen. Upper-limb prostheses are artificial devices that assist users from being restricted to their motor deficits and to achieve motor tasks smoothly. Functional upper-limb prosthetic training can improve users' performance in controlling prostheses and has been incorporated into occupational therapy for individuals in need. In recent years, virtual reality (VR) and augmented reality (AR) technologies have been shown to be promising avenues for improving the security, convenience, and efficacy of rehabilitative prosthesis training systems. Given the widespread use and success of virtual and augmented reality systems, many neurorehabilitation developers have begun incorporating these paradigms into upper-limb therapeutic services with promising results. However, it is still unclear what advantages these paradigms have over one another when applied to prosthesis control. What's more, it is uncertain whether or not the comprehensive efficacy and effectiveness of VR or AR assistive tools are adequate compared to conventional prosthetic tools and if not, whether the gap can be narrowed through incorporation of other technical paradigms.

Aiming to find the answers to these problems, this work first presents a mixed reality system for prosthesis control and training called VivePHAM which is a simulation of a real-world functional outcome assessment system—Prosthetic Hand Assessment Measurement (PHAM). The VivePHAM can provide users with virtual prosthesis control and training in both VR (VRPHAM) and AR (ARPHAM) environments. We recruited a population (N=5) of able-bodied subjects and ask them to perform three-dimensional object manipulation tasks in analogous AR and VR environments respectively. Evaluation metrics such as path efficiency, task completion rate and time, and Fitts' throughput are applied to assess subjects' performances within the two paradigms. Based on the comparative analysis of these data from the study, we find that VR-based environment promotes more efficient motion along with higher task completion rate and path efficiency while AR paradigm allows subjects to perform motor tasks with shorter time consumed. Another study is conducted to evaluate the efficiency and feasibility of AR-facilitated prosthesis control system compared to that in real-world and if any technical additions can be applied to improve AR-based system. A population of (N=3) able-bodied subjects were engaged in the experiment to perform object manipulation tasks in physical environment, AR-without-bypass environment and AR-with-bypass environment respectively. Based on the results obtained from the assessment of evaluation metrics such as peak-velocity, path efficiency, trajectory range and time consumed, we find that while our AR-based system modestly lags behind the effectiveness of that in physical system, the equipment of a bypass prosthesis can improve the efficacy of AR system in prosthesis control.

**Primary Reader:** Nitish V. Thakor, Ph.D.

**Secondary Reader:** Scott M. Paul, Ph.D.

**Secondary Reader:** Peter Kazanzides, Ph.D.

# Acknowledgements

First, I'd like to give most thanks to my principal investigator: Dr. Nitish V. Thakor who provides me with the valuable opportunity to conduct my research studies in the Neuroengineering and Biomedical Instrumentation Lab (NBIL). Dr. Thakor gives instructive guidance and sufficient support during my research endeavor, which help keep my work pace smoothly and firmly and without which, this thesis would not come into fruition. Next, I'd give most sincere thanks to Christopher L. Hunt. As a sagacious senior PhD candidate student and a bosom friend, he led me into the realm of research from a bachelor, helped me overcome many difficulties and walked me through my most struggling time like an elder brother. Special thanks to Dr. Alcimar Soares who gave me detailed instructions on my work at early stages and memorable collaboration experiences with teams at Federal University of Uberlândia. Another special thanks to Dr. Rahul Kaliki whose supervision keep my work on track and who provides me with sufficient resource support. Thanks to Dr. Jacqueline Hebert's team at the University of Alberta who provided much technical support and information. Thanks to my amazing colleagues Mark Iskarous, Sriramana Sankar, Rebecca Greene, Catherine Ding and Arik Slepyan in our lab who together form a friendly and nurturing research environment to work in and bring me memorable experiences.

# Dedication

This thesis is dedicated to my loved parents and grandparents who raised me up and always love me and support me without any reservation nor hesitation. They are always the warmest harbor no matter where I might unpredictably drift into. Also dedicated to all warm-hearted people I met, it is you that bring color to my life and illuminate this world.

# Table of Contents

# List of tables

# List of Figures

# Chapter 1. Introduction

This chapter gives background introduction about the status of neuromotor disorders and current treatment methodologies, and a brief introduction about the development of virtual reality and augmented reality and overview of their application status in rehabilitation area. We then elicit a needs analysis in treatment of neuromotor disorders that result in upper-limb motor deficit, with specific focus on post-stroke treatment and prosthesis-based treatment, and how the incorporation of VR or AR technology can fulfill these needs properly. This work is quoted from part of my work on the review chapter: Neurorehabilitation with Virtual and Augmented Reality Tools included in an upcoming book publication: Handbook of Neuroengineering.

## 1.1 Neuromotor disorders and conventional treatment methods

Neurorehabilitation aims to facilitate the recovery of individuals suffering from neurological disorders that cause physical impairments and disabilities, such as stroke and traumatic brain injuries. Major goals of neurorehabilitation technologies include alleviating symptoms, restoring functionality and mobility, and improving patients' independence towards activities of daily living (ADL). Stroke, also known as cerebrovascular accident (CVA), is a major disease leading to long-term adult disability worldwide [1] as well as the most common disease resulting in dexterity impairment of the upper limbs [2]. Studies show that one out of six individuals in the world will have a stroke over the course of their life [3], with 795,000 reported stroke episodes occurring annually in just the United States [4]. Many people who have survived a stroke still suffer from minor to severe cognitive or motor impairments, such as limited range of motion (ROM), muscle weakness, and problems with balance and coordination [5]. These post-stroke impairments result in a negative impact on a patient's autonomy, by limiting their ability to perform self-care tasks as well as communicate with others effectively [5][6].

Strategies for stroke rehabilitation aim to reverse the resulting functional deficits through regular motor and cognitive exercise, with the end goal being an increase in performance of common ADLs. For motor impairments specifically, rehabilitation regimens tend to focus on increasing the strength of voluntary muscle activity [7][8]. To accomplish this, post-stroke rehabilitation strategies generally ask patients to perform repetitive motor tasks that require active cognitive attention to take advantage of post-traumatic neuronal plasticity and induce functional reorganization in the motor cortices [9][10]. Unfortunately, to be effective, these strategies often require long periods of cognitively demanding, monotonous motor task performance. Because of their repetitive nature, patients undergoing conventional therapies find maintaining motivation difficult, resulting in limited effectiveness of the intervention [2][11]. For best outcomes, rehabilitation must be repetitive, intensive, and precocious. The challenge then is to design a rehabilitation format that maintains patient motivation and dedication [12].

Another major type of neuromotor deficit, limb loss can have a profound effect on various aspects of life. According to Ziegler-Graham et al., in 2005, around 1.6 million individuals were living with limb loss in the United States. Projections from the same study show that the amputee population is likely to expand more than twice as much by 2050 [13]. Nowadays, it is estimated that around 2.2 million persons suffer from limb loss, among which more than 60,000 have significant upper-limb loss [13]. According to the statistics from National Limb Loss Resource Center®, there are approximately 185,000 amputations that take place in the United States each year [14]. It is undeniable that any limb loss will bring considerable impacts to amputees' daily life. Amputees will not only encounter functional challenges, such as dramatically reduced locomotion and fine motor skills, but amputation may also have adverse effects on individual independence and psychosocial expressions. As a dexterous and irreplaceable part of the body, our hands assist us in capturing useful information and expressing our thoughts and emotions intuitively. As such, the loss of a hand will inevitably contribute to a reduction of ability to interact with the surrounding environment and perform various social skills [15][16]. Besides the functional and social inconveniences, there are potential sequelae after amputation surgery that may bring a sense of both physical and mental discomfort to amputees. For instance, it is estimated that 60% to 80% of individuals with limb loss experience phantom limb sensations, the subjective feeling and sensation of the unrealistic existence of the absent limb after traumatic injuries [17][18]. Moreover, the incompleteness of the physical body's appearance might arouse passive negative emotions or lay spiritual pressure on survivors with limb loss. A study applying a standardized assessment procedure shows a depression rate of around 35% in an outpatient population with lower limb amputations [19].

While the factors contributing to limb loss vary from person to person, the major causes leading to amputation around the world are trauma, disease, and congenital limb deficiency, in order of incidence [20]. In most developing countries, trauma caused by motorized vehicles and machinery and unproperly treated bone fractures are the leading causes of amputations, causing ~80% of all amputations. In contrast, vascular complications caused by diabetes lead to most amputations in developed countries [20]. In the United States, vascular diseases and trauma account for 54% and 45% of all limb loss cases, respectively [1].

One of the most efficient and well-adapted methods to assist individuals with limb loss with mobility impairments is to equip them with prosthetic limbs. In recent decades, prostheses have evolved from an artificially-crafted work done by artisans into a technically intelligent and dexterous engineering device. Burgeoned by advances in sensors, computing hardware, and machine learning techniques in recent years, modern neuroprosthetic limbs have been greatly improved with respect to controllability, functionality and so on, making it more adoptable for use in daily life. In particular, myoelectric prostheses have been widely adopted for clinical usage for upper-limb amputees. By monitoring muscle group activations through surface electromyography (sEMG), a myoelectric prosthesis can help users accomplish complex manipulations and motor control. Prosthesis control is achieved through different patterns of contractions generated by the user's residual muscle groups. Electrodes placed on the residual limb's surface sample the sEMG signals from the residual muscle groups, which are then interpreted through control paradigms, such as direct control or pattern recognition, into specific classifications of motor activities [21].

In recent years, pre-fitting training has become an essential part of neuroprosthesis treatment even before amputees gain their physical prosthetic device. Current training methodologies for myoelectric control are quite labor-intensive, requiring the involvement of occupational therapists and other medical staff. However, current methodologies hardly meet the expectations of their users in terms of effectiveness. In order to gain better control of their prosthetic device, traditional pre-prosthetic training protocols require amputees to actively practice eliciting residual muscle activity with guidance from an occupational therapist. Through pre-fitting training, amputees are to learn how to contract different residual muscle groups with enough specificity and distinguishability to coordinate prosthesis functions, such as hand open or hand close. However, there is a lack of intuitive feedback for amputees to correct or adjust their muscle control during this process. Moreover, considering the significant period of time between the amputation surgery and the arrival of the customized prosthesis, the training outcomes with the temporary prosthesis will be largely discounted by the time they are applicable. More physical and mental effort is needed while amputees are adapting to their customized devices. New fitting issues may come up, furthermore burdening users and therapists, leading to an ineffective training experience [22][23][24]. Such factors partially contribute to the large number of amputees who later reject or abandon their prosthetic devices. In fact, some of the upper-limb amputees may instead gradually get accustomed to managing their daily-encountered tasks with a single hand and find it more cumbersome to operate with their customized prosthesis [25]. A survey in 2011 that focused on the population of upper-limb loss users found that the primary reasons for rejection of prosthetic devices—especially secondary prosthesis rejection, come from the dissatisfaction with prosthetic comfort, function, and control [26][27].

A good myoelectric prosthesis training system should meet requirements from various parties, including clinicians, therapists, and amputee users. For example, a system should be financially accessible to the userbase that it serves so as to not economically burden its users. Furthermore, training devices should be portable and easy to carry along so as to reduce logistical barriers to protocols that require frequent use [28]. In addition, it would be ideal and convenient for users if they could train with the system by themselves remotely, without restrictions on time or place, allowing users to self-pace their learning. Moreover, in order to ensure the efficacy of the overall rehabilitation process, if would be best if the training protocol to maximally mimic the circumstances in which the prosthesis will be used after having access to it [22].

## 1.2 Virtual and Augmented Reality and Rehabilitation

In recent decades, virtual reality (VR) has become a trending immersive technology with applications in a wide variety of different disciplines [29]. VR technology has been a public point of attention since the 1980s, with the first commercial VR system designed for VPL research company (San Francisco, CA, the U.S.) [30][31]. Since then, VR technology has been dramatically improved due to advances in both hardware and software, such as the enhancement of graphic resolution due to the evolution of GPUs and processors, and more immersive visual experience with the development of headset [32]. Today, VR suggests completely immersive experience in an artificial visualization that shuts out the physical world. An extension of VR, augmented reality (AR), takes these virtual visualizations and portrays them overtop the existing physical reality. In more advanced systems, virtual elements may interact with the physical context they are displayed in; we refer to these advanced experiences as mixed reality (MR).

Following the maturity of many VR and AR technologies, a broad range of industries, such as education, energy, astronomy, and defense, have begun applying these immersive environments to resolve emerging problems and facilitate productivity [33]. Healthcare and medical fields have also begun to take advantage of such technologies. Comprised of VR or AR to provide users with enhanced visual experiences and other assistive technologies to make it better fit for rehabilitation use, VR/AR assistive tools have been applied for rehabilitation purposes for many years (Figure 1).

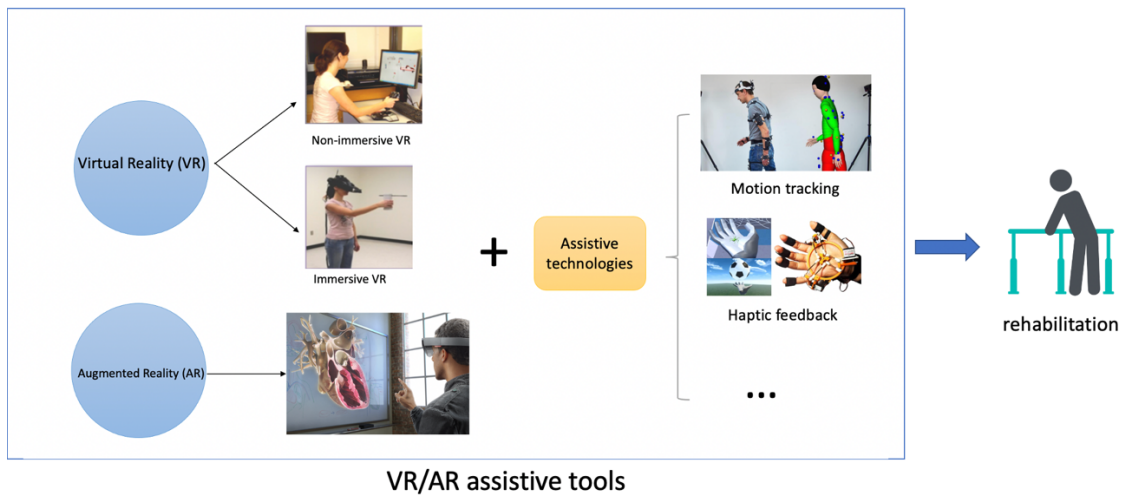

**Figure 1. The technical structure of VR/AR assistive tools for rehabilitation. A VR/AR assistive tool typically applies non-immersive or immersive virtual reality or augmented reality to form a virtual environment for users to conduct rehabilitation activities, along with other assistive technologies such as motion tracking, haptic feedback and so on for enhancement of performances.**

As far as neuromotor impairments are concerned, the intervention of VR and AR assistive tools has been proposed and well-applied to a variety of therapies for a range of disorders, such as stroke, cerebral palsy, and Parkinson's disease. These tools have found success through their ability to maintain patient motivation with entertaining, gamified rehabilitation protocols [34]. Furthermore, in recent years, researchers and developers in the rehabilitation field have also started to apply VR and AR technologies to pre-fitting prosthesis training. Early examples include the MyoBoy® software suite (Ottobock HealthCare, Duderstadt, Germany) which aims at guiding users in myoelectric prosthesis training with sEMG biofeedback [35]. MyoBoy® can display the sEMG signals collected from electrodes along with a visual representation of the hand on the computer screen; it also contains a simple interactive game for users to undergo prosthesis training wherein users control two cars whose position is determined by the amplitude of the sEMG signals. Users must guide these cars through obstacles that periodically appear on the screen. With this interface, users train their residual muscle groups to eventually control and switch between various hand postures, such as hand open/close or wrist pronation/supination (Figure 2).

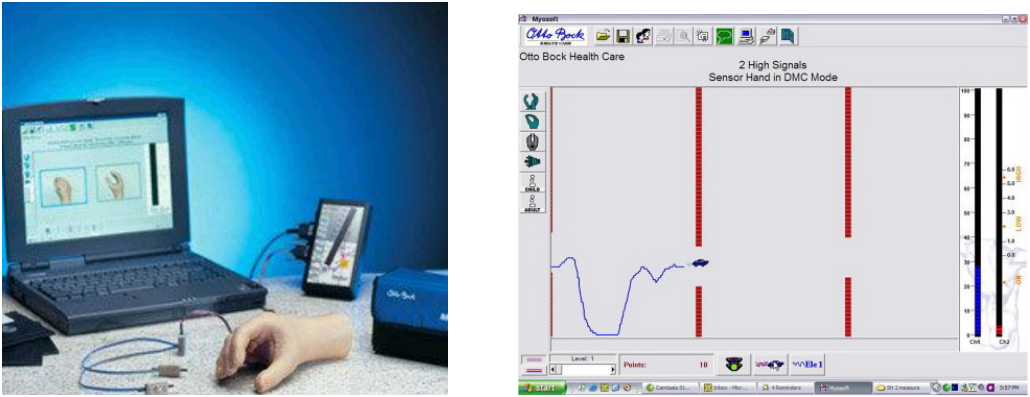

**Figure 2. Prosthetic training through MyoBoy® software game (Ottobock HealthCare, Duderstadt, Germany). In the game interface, the trajectory of the car represents the fluctuation of the surface EMG signal. Users will need to control the movement of the car to pass through the gates as indicated via myoelectric control. The game aims to improve users' mastery in controlling the myoelectric input with higher accuracy and endurability [24]. Reprinted with permission.**

And nowadays, some VR/AR devive companies such as HTC, Oculus, and Microsoft, have developed commercial VR and AR devices, such as head mounted display devices and AR glasses that are technically mature, performant, and affordable for the average consumer, which facilitate the application of VR and AR technology in rehabilitation areas (Figure 3) [36][37].The incorporation of VR and AR paradigms provide users with an immersive virtual rehabilitation environment in order to improve users' training experience and reduce the burden on both the patients and clinicians involved [38]. These environments are aided by real-time visual feedback of the prosthesis' controllability without requiring the user to suffer from the discomfort brought on by premature mechanical interactions when the surgical trauma of the residual limb is not completely healed [39]. Neurorehabilitation in VR and AR environments can provide a flexible and customizable experience since modifying the virtual components of a rehabilitation environment is much easier than doing the same for traditional, physical environments [32]. Furthermore, VR or AR-based rehabilitation modalities are also simpler to set up, more portable, and more financially affordable when compared to traditional, clinical methods [40].

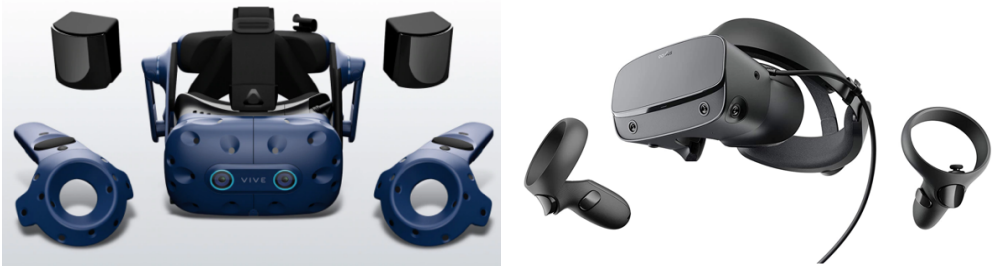

Figure 3. Two popular, commercial VR headsets: Vive ProTM Headset by HTC Corporation© (left) and Oculus Rift STM VR Headset by Facebook© (right). Current VR head mounted display devices can provide users with immersive visual experience in virtual or augmented reality [36][37]. Reprinted with permission.

Conventional rehabilitation methods for neuromotor disorders lay heavy emphasis on repetitive therapeutic exercises or training to counteract functional limitations and restore mobility. However, these conventional therapies often result in decreased patient motivation, leading to non-optimal outcomes. This has led to a concerted effort to design novel rehabilitation methods that are more patient-friendly and engaging, resulting in modalities for functional restoration to evolve dramatically in recent years [41]**Error! Reference source not found.**. However, there are still many needs and challenges to address in order to enhance the effectiveness of neurorehabilitation therapies. In the following section, we analyze the current needs in modern rehabilitation, especially those in post-stroke and neuroprosthesis treatment, and illustrate how the incorporation of VR and AR technologies can be used to help address those needs and promote the effectiveness of those treatments.

## 1.3 Needs effectiveness for post-stroke treatment

Post-stroke rehabilitation therapies aim to exploit the plasticity of the nervous system to allow patients to recover functionality and alleviate motor deficits. In fact, the most extensive and spontaneous post-stroke improvements occur in the first 30-90 days of recovery due to heightened neural reorganization [42]. Therefore, the effectiveness of a post-stroke treatment is a direct result of the quality of the intervention during this period. As mentioned previously, conventional post-stroke therapies require patients to repetitiously perform sequential motor tasks in an effort to strengthen the voluntary control of weakened motor units [43]. This form of therapy lends itself to patient disinterest over time due to it being both physically and cognitively demanding with little variation. Furthermore, requiring patients to perform these therapies in a clinical setting results in added detrimental factors, such as limited treatment resources from a clinical institution and general transportation costs for mobility-limited individuals [44]. Therefore, a novel treatment paradigm that engages users and retains motivation during the first one to three months of rehabilitation while simultaneously providing patients with easier access to periodic treatment is expected to address the needs.

VR has been applied as a tool in rehabilitation since the 1990s and has bolstered recovery efforts for patients with a variety of neurological disorders [45]. For example, an overview by Lange et al. revealed the growing popularity of VR game usage in movement rehabilitation and chronic neurological treatment over the past decade [46]. One of the benefits that VR as a treatment intervention tool offers is an increase in user motivation and engagement during the therapeutic program, which consequently positively impacts treatment outcomes. Another desirable feature of VR is that tasks and scenarios can be easily customized based on user needs since there is no need for physical modifications to the rehabilitation protocol. Furthermore, VR environments provide users with a safe space to practice potentially risky motor activity necessary for ADLs. An example of this is practicing street crossing in VR to help a patient gain confidence in their mobile autonomy in a safe and controlled environment [47]. Finally, VR as a paradigm offers post-stroke patients a more convenient location to conduct their regimen. Instead of transporting themselves to a clinical institute for rehabilitation, patients can practice directly at home, assisted by portable VR devices.

## 1.4 Needs effectiveness for neuroprosthesis treatment

The pre-fitting period is an indispensable part of neuroprosthesis treatment in which the subjective training experience directly determine a user's chronic adaptation to their prosthesis and its eventual adoption or abandonment [48]. Therefore, efficacy during this phase represents one of the major challenges to improving the usability of prosthetic devices and user satisfaction.

As mentioned earlier in this chapter, myoelectric-based prosthetic devices can be considered a breakthrough when it comes to helping the target population restore lost motor function and mobility after amputation. However, myoelectric-based prostheses often require a considerable amount of training effort, which can be amplified depending on the desired functionality and control strategy. The more sophisticated the prosthesis, the more distinguishable residual muscular activity must be to generate the different myoelectric patterns required to control the prosthesis' various degrees of freedom (Figure 4) [16][49]. Prior to the arrival of a customized prosthesis, users learn to intentionally activate or contract their muscle groups and isolate specific myoelectric patterns. Such training is usually performed under the guidance of occupational therapists in rehabilitation institutions [50]. In spite of the fact that amputees will make progress in the long-term training and gradually learn to activate their residual muscles for prosthesis control with less cognitive effort, many find the overall process too cumbersome or exhausting and abandon their prosthetic device before breakthroughs in training are achieved [51].

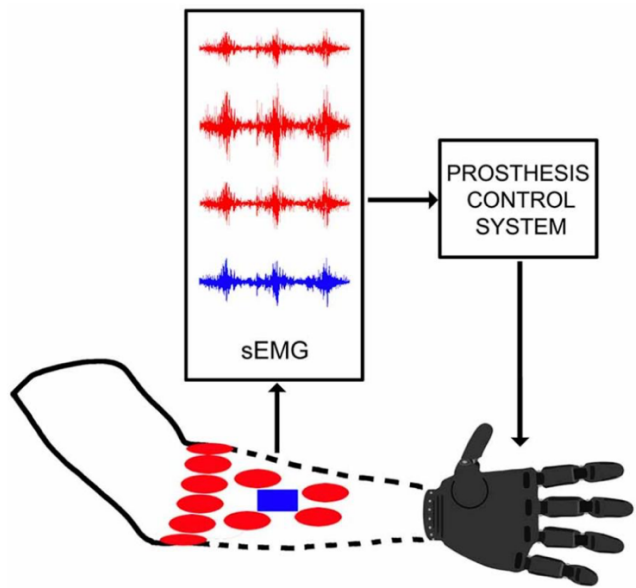

**Figure 4. The system architecture of a general myoelectric prosthetic system. Surface electrodes are attached on the muscle groups of the residual limbs to sample the surface EMG signals which are then decoded to control the motor activities of prostheses [49]. Reprinted with permission.**

Another major issue with conventional pre-fitting prosthesis training is the inaccessibility of feedback during myoelectric control practice. Such lack of feedback prevents users from intuitively acknowledging the correspondence between applicable motor activities on prosthetic hand and specific muscle contractions until they can sense the activities of their physical prosthesis or are explicitly guided by their therapist [24].

Ownership and proprioceptive awareness of the artificial limb are considered the most significant missing features in current prosthetic devices [27]. In recent decades, researchers have facilitated the adoption of prosthetic devices by trying to enhance the perception of the prosthesis as a part of users' body[52]. The formation and congruency of visual, proprioceptive, and behavioral cues towards a prosthesis are paramount for a sense of embodiment [53]. Therefore, it is necessary for current pre-fitting protocols to provide various modalities of feedback (beyond just visual) so that users may gain a sense of embodiment over the prosthesis during that preliminary phase. Technical paradigms that accommodate such needs help dissipate concerns amputees may have regarding their real devices' usefulness.

Additionally, data generated during sEMG or EEG prostheses training, such as the performance of myoelectric pattern recognition (MPR) classifiers or the energy consumed by the residual musculature to perform a particular motor task, may convey meaningful and instructive information to therapists and trainees [54][55]. Such information plays a significant role in guiding therapists towards personalizing the training plans and schedules that would fit the individual rehabilitative needs of amputees. However, there are hardly any technological tools involved in conventional training methods that can visualize and replay the data collected during training to trainees and occupational therapists in real-time [16].

With the advancement of VR/AR technology, the capacity of these techniques to provide users with an immersive virtual pre-fitting training model has improved at a swift pace. VR-based training interfaces have already shown promising results assisting therapists and amputees during the pre-prosthetic training phase [56]. The essential principle of VR-based training interfaces is to provide amputee users with a projected virtual prosthesis attached to their residual limb. During the training process, the sEMG signals generated by the residual musculature will drive the virtual prosthesis' motor activity, which is then displayed to the user in real-time. Based on such rationale, some studies have been focusing on the development and optimization of classification algorithms for signal processing and decoding for better conversion for virtual prosthesis control. A study by Betthauser et al. focused on the development and validation of an EMG sequential model with temporal convolutional networks that can achieve outstanding prediction of movement [57]. With simpler feature decoding and shorter response delay (4.6 ms), this model can yield predictions that is more accurate and stable (p<0.001) compared to other frame-wise models in this field, particularly regarding inter-class transitions (Figure 5).

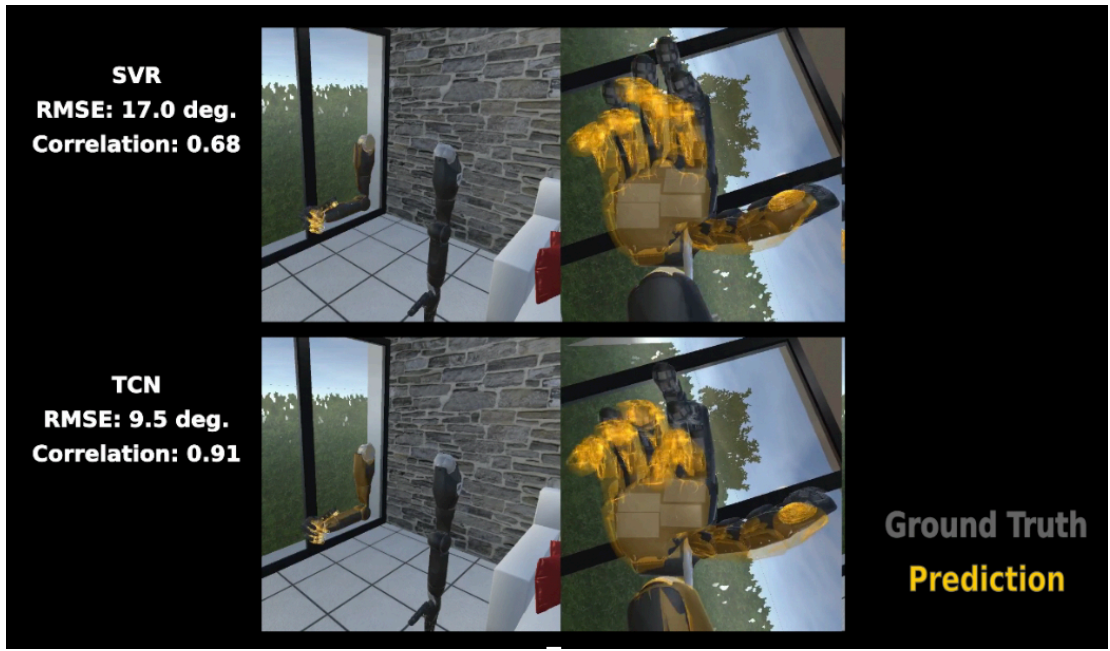

Figure 5. Validation display of movement predication with temporal convolutional network model for EMG signal processing. The virtual reality environment displays the prediction of the prosthesis movement in real-time (colored by yellow) predicted by the model based on the EMG input [57]. Reprinted with permission.

Another example include the Augmented Reality Myoelectric (ARM) Trainer developed by Anderson et al., from which users could see the virtual arm overlaid on top of their residual limb and could control the virtual prosthesis by activating their muscles in the same way they would control a physical prosthesis (Figure 6) [22]. In contrast to conventional prosthetic training procedures, in which users would have to voluntarily activate their muscles without any feedback, the level of user interaction and active participation is dramatically enhanced through real-time visual feedback provided through XR. Game-based or immersive video-based prosthetic training interfaces that leverage XR technology may increase user motivation towards training by reducing the monotony of traditional training protocols. For example, several studies investigating the Nintendo Wii as a rehabilitative therapy tool have shown that by increasing the user's entertainment during therapy, their motivation towards accomplishing the necessary therapeutic tasks increases as well [58][59].

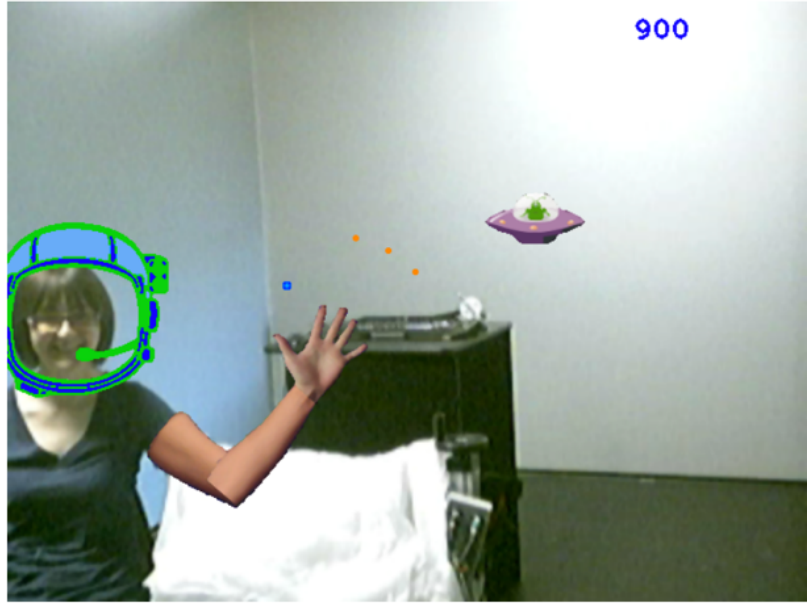

**Figure 6. The ARM Trainer prosthetic training system. The system aims to train users on controlling muscle group activities intentionally for myoelectric prosthesis adaption. Users can see a mirrored view of themselves with virtual arm superimposed on their residual limb. Users are able to control the motor activities of the virtual arm through muscle contraction to interact with the interface and achieve the tasks in the game [22]. Reprinted with the permission.**

While providing immediate control feedback to the user, the visual cues of a virtual prosthesis attached to the user's physical body can also provide users with a feeling of ownership towards their prosthesis, helping amputees integrate a representation of the prosthesis into their body image before using a physical device. Furthermore, incorporating VR/AR interfaces into the rehabilitation process can improve basic balance control and mobility [60][61]. A study by Sveistrup et al. compared the treatments of groups of patients with post-traumatic brain injuries treated with and without the intervention of VR technology. The results show that though both of the methods improve patients' balance, the group treated by VR-based therapy was found to provide patients with the additional benefits of increased patient motivation, increased patient confidence during walking, and decreased patient fall likelihood [62]. Another benefit of VR/AR-based rehabilitation is the flexibility to customize the training formats and individualize the specific rehabilitative motor tasks for different users depending on their specific needs for function restoration or improvement. Since VR/AR applications can be programmed dynamically, therapists and amputee users can make online adjustments to the training tasks and modules, increasing repetitions of a particular type of practice based on the performance during the training [63][64]. Furthermore, informative data regarding the kinematic movement and the trainees' quantitative performance can be recorded by most VR/AR-based training systems to provide occupational therapists and prosthetists with important information reflecting the trainees' performance during training. For example, a computer-assisted rehabilitation environment called Computer Assisted Rehabilitation Environment (CAREN) developed by Motek Medical can record joint ranges of motion (ROM) and movement symmetry for later assessment and analysis [65][66].

XR technologies can also be applied to post-amputation therapy to help alleviate phantom limb pain, having achieved remarkable performance in clinical treatment [67]. Traditional methods for treating post-surgery phantom limb pain involve, for instance, the use of mirrors showing the intact limb of patients instead of the amputated one. This type of mirror therapy provides the visual illusion of having two intact limbs extended from both arms [68]. Inspired by such methods, an immersive VR system can simulate an intact limb projected onto the patient's residual limb and virtualize the limb's movement by tracking the users position and orientation. As a result, patients can gain a virtual representation of their body with a simulated intact limb in real-time through a visualization via a head-mounted display device (Figure 7) [69][70].

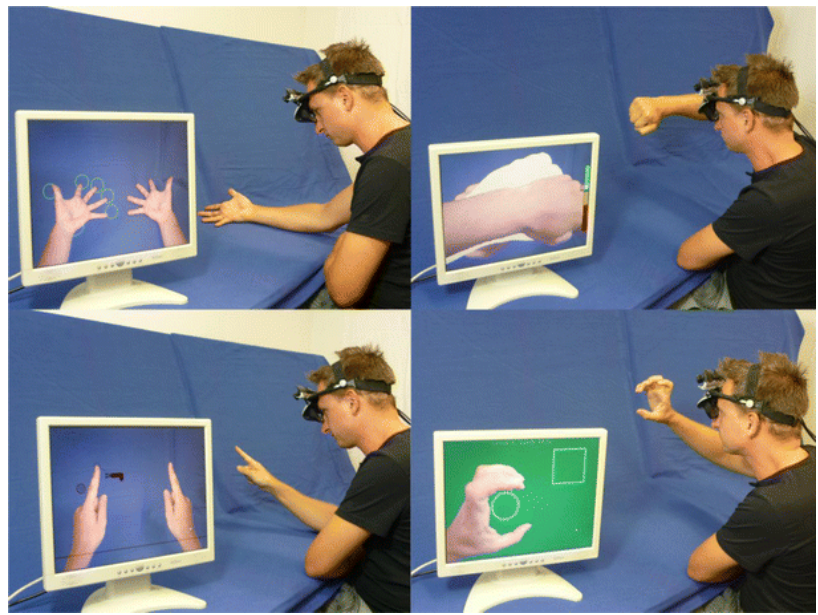

Figure 7. Augmented reality training system for phantom limb pain treatment. The system developed by Jörg Trojan et al. is based on the principles of mirror therapy [70]. The system aims to aid the treatment of phantom limb pain, hemiparesis after stroke and so on. Users will go through a series of tasks indicated in the figure and they can see the mirrored image of their hand through head mounted display. Reprinted with permission.

# Chapter 2. Review of Virtual and Augmented Reality-based Applications for Neuromotor Disorders

In this chapter, we give a comprehensive review on some representative VR and AR assistive tools and cutting-edge applications applied in neurorehabilitation area. We analyze how these tools utilize the virtual technologies to enhance the treatment of these disorders compared with conventional therapies, and further discuss about the advantages, salient features along with the limitations on these VR and AR applications. The content of this chapter is partially quoted from my work on the review chapter: Neurorehabilitation with Virtual and Augmented Reality Tools included in an upcoming book publication: Handbook of Neuroengineering.

# 2.1 An Overview of Virtual and Augmented Reality Applications in Neurorehabilitation

This section introduces and analyzes state-of-the-art VR/AR applications in neurorehabilitation and discusses the pros and cons of these applications compared to conventional treatment procedures. The effects on treatment outcomes, the advantages and benefits VR/AR can potentially bring to patients and clinicians, as well as the current limitations are explored.

## 2.1.1 Applications of virtual and augmented reality in post-stroke therapy

As mentioned earlier in the previous chapter, the level of engagement and motivation during a therapy program is crucial to the final outcome of a chronic treatment. In virtual reality, serious games have emerged as an advantageous and enjoyable alternative to traditional therapies [80][81]. A serious game is a game that goes beyond pure entertainment, as in video games, since it offers other types of experiences that can be directed to relearning, rehabilitation, and training [82]. Due to the limitations on computer graphics and visual rendering in early decades, traditional virtual games were limited to simple 2D environments or low-quality 3D environments, neither of which elicit a significantly positive impact on increasing participants' interest in their rehabilitation. Moreover, in many of these games, it is not possible to adapt the rehabilitation tasks, according to the patient's clinical stage[83].

Recently, Cyrino et. al. proposed a customizable serious game, called HarpyGame, which is based on non-immersive VR techniques. The game was designed to achieve a more natural and intuitive interface for post-stroke rehabilitation [84]. It is well known that the sooner the rehabilitation process starts for a post-stroke patient, the faster they will recover most (or all) of their motor abilities. To achieve this, the game consists of a 3D environment where the patient controls a harpy eagle, a typical rain forest bird, using the positioning of their arm. The virtual environment consists of a virtual forest containing a series of objectives and challenges, such as flying through a specific route, avoiding obstacles, and catching food. Compared to conventional therapy, such a game in VR can enhance the motivation of participants, by being visually stimulating and allowing adaptive exercises that coordinate with patients' interests and habits. Figure 8 presents examples of the game interface.

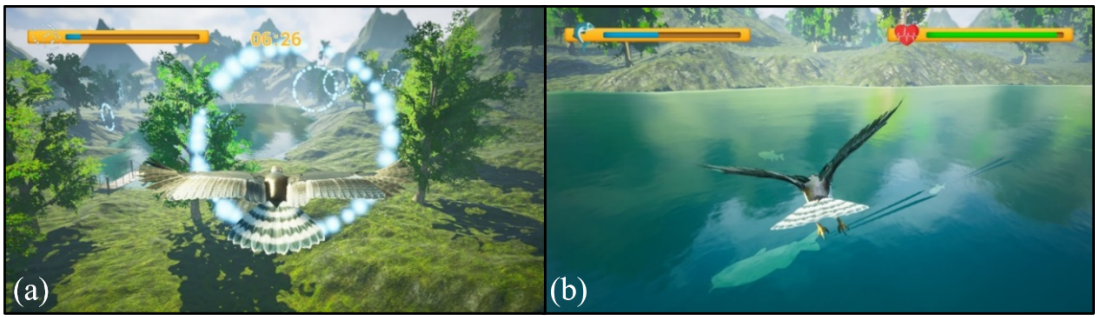

**Figure 8. Screenshots of the HarpyGame. (a) The user controls the harpy eagle in the virtual game by using the positioning of their arm and pass it through the cloudy circles. Figure 8(b) shows the user controlling the eagle to catch a fish. In doing so, a patient can recover their motor abilities by means of entertainment. All movements follow physiotherapeutic protocols [84]. Reprinted with permission.**

Through playing the HarpyGame, the patient ends up performing the following exercises: internal and external rotation of the arm enables movement along the right-left axis, while elbow flexion and extension moves the character along the up-down axis. After the execution of a session, it is possible to generate a graph containing the trajectory curve traveled by the harpy eagle, which can be used to analyze the entire patient's progress in the scenario. Using these trajectories, therapists can evaluate patient performance of a motor task by comparing the actual trajectory traveled to the anticipated path designed by the game and calculate the accuracy of the motion.

Another advantage of the HarpyGame is that a therapist can adjust and configure the level of challenge according to the patient needs or clinical conditions. Furthermore, the game supports multimodal interfaces, such as sEMG armbands and direct motion capture, which make it customizable for each patient with different physical capabilities post-stroke. Moreover, depending on the patient conditions, the treatment can start even before the patient is discharged from the hospital. Figure 9 shows two sets of input devices supported by the game: the Myo® device and an X-Platform.

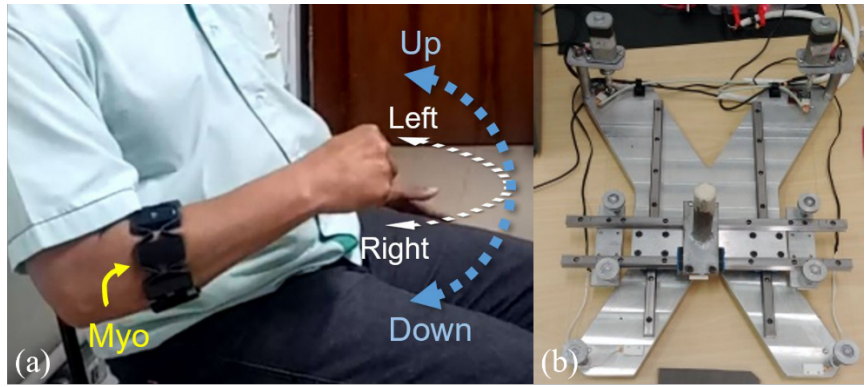

**Figure 9. Different input devices for the HarpyGame. (a) A stroke patient using the Myo® device. The forearm positioning data is captured and sent to the game. Data from the device's accelerometer translates the horizontal and vertical accelerations into the harpy eagle movements. (b) An X-platform to capture user's movements. This platform has a mechanism composed of motors driven by cables that act in a Cartesian plane. In this mechanism, the patient performs a movement by means of a rod attached to a rail connected to the cables, which drives the motors, providing assistance feedback. The therapist can configure this assistance to generate an elastic force on the patient's arm, assisting him in executing the movement, or restricting the movements requiring muscle activation [84]. Reprinted with permission.**

Another study by Assis et al. was conducted to assess the efficacy of AR-based training for post-stroke recovery [85]. An AR training system called NeuroR was developed and applied in the study (Figure 10). The system used an RGB camera in conjunction with an instrumented glove to track the motion of a patient's injured arm in real-time and project a simulated arm into the augmented scene. Four subjects diagnosed with stroke participated in the study. The NeuroR system relies on motion tracking and a physiotherapist to instruct the patient to complete task-oriented exercises with the virtual arm. In clinical testing, the experiment applied the Fugl-Meyer scale, variance on ranges of motion before and after the training to assess a participant's upper-limb functional outcomes [86]. The results showed an increase in range of motion both in the AR training group (46.7% to 73.9%, +27.2%) and in the conventional group (61.3% to 90%, +28.7%). This statistically significant increase on the Fugl-Leyer scale for the AR training group ($p<0.05$), demonstrates the efficacy of AR systems like NeuroR in assisting post-stroke treatment.

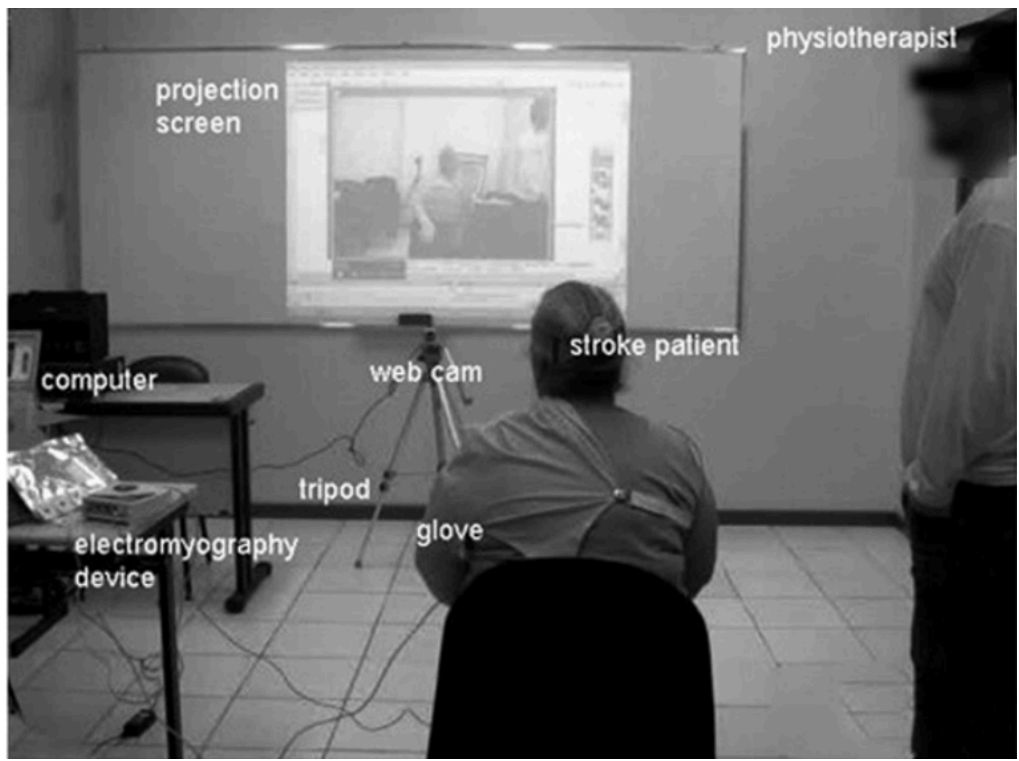

**Figure 10. NeuroR AR system. A web camera captures the patient's body image in real-time, which is then projected onto the screen in front of the patient. A glove containing motion trackers is worn on the patient's hemiplegic arm to track movement. When the system starts, a virtual arm is casted near the injured arm of the patient in the screen. Patients are**

**then prompted to move their injured arm to overlap the virtual arm during the training [84]. Reprinted with permission.**

As for immersive VR rehabilitation environments, AlMousa et al. developed an immersive VR-based gaming system called Move-IT for upper limb stroke recovery training [87]. The system uses the Oculus Rift head-mounted display to provide visual feedback and utilizes a Leap Motion hand tracker to get inputs of hand motion from the user. As shown in Figure 11, the scene in the game is displayed in a first-person point of view for immersive interaction experiences. The game revolves around the player (the patient) who need to sort colored cubes. The game's objective is for the player to sort all the cubes into the correct colored bins. Move-IT also incorporates a timer and a scoring system. The timer is used to set an appropriate level of difficulty for the user as well as to make the game more engaging. It counts down from a target time each level, and the player attempts to sort all the cubes before the timer reaches zero. Each level's target time is customized over multiple iterations for a specific user by a physical rehabilitation therapist. This is done so that the target time can be reasonably achieved, but the game still poses a fun challenge. The purpose of the scoring system is to give the user a positive feedback when they perform well, and points are awarded for each cube sorted into the correct bin. The game offers four levels to promote engagement, and each level involves performing a variety of upper extremity rehabilitation exercises including shoulder flexion, extension, abduction, adduction, as well as hand extension, grasp, and release. Five participants with stroke played the game over multiple sessions with a physical therapist. By evaluating the time consumed, the performance errors and required assistance, as well as a follow-up questionnaire investigating participants' subjective feelings towards the training process as assessment metrics, the study showed the feasibility of the immersive VR gaming system for upper limb stroke rehabilitation.

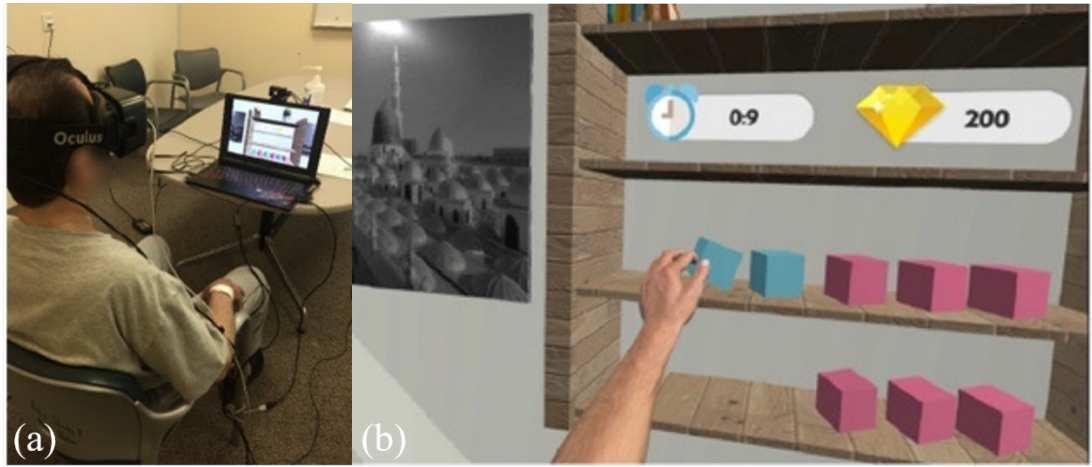

**Figure 11. An overview of the Move-IT system. The objective of the game is to sort and relocate cubes to designated boxes with corresponding colors. There are shelves full of red and blue colored cubes in front of the player, as well as correspondingly colored bins to their left and right. Users need to control the virtual arm in the game by moving their upper limb to achieve these relocation tasks within the set time [87]. Reprinted with permission.**

## 2.1.2 Applications of virtual and augmented reality in chronic neck pain therapy

Neck pain is another neurological disorder, and it is reported that more than 14.4% of adults in the United States have experienced neck pain. Especially with chronic neck pain, individuals' health, comfort, and productivity are negatively affected [88].

Even though there are some interventions to relieve chronic neck pain such as pharmaceuticals, injection, and some neck rehabilitation exercises like stretching [89], there are still many patients who experience discomfort and disability during and after rehabilitation; many of them may have developed kinesiophobia that is a psychological pain-related fear for movement and physical activity [90], which may delay the rehabilitation and recovery. In recent years, virtual reality (VR) has been widely used in the field of rehabilitation engineering as technologies continue to advance. VR offers a simulated environment that contains computer generated graphics. Using VR, patients stay more engaged and motivated in the exercises through the use of game-like experience in VR [91]. With the help of VR, the patient outcomes are likely to be improved and costs may also be reduced with reduced frequency of clinical visits. In the scenario of treating or relieving chronic pain especially in patients who have developed kinesiophobia, VR is a viable tool to reduce their fear as the gaming interface and designed computer graphics lead to distractions and potentially the patients can gradually conquer the pain and recover from what they have suffered for years. Chen et al. examined the use of altered visual feedback from virtual reality for patients with chronic neck pain and kinesiophobia and demonstrated that the virtual reality is useful in the rehabilitation process [92].

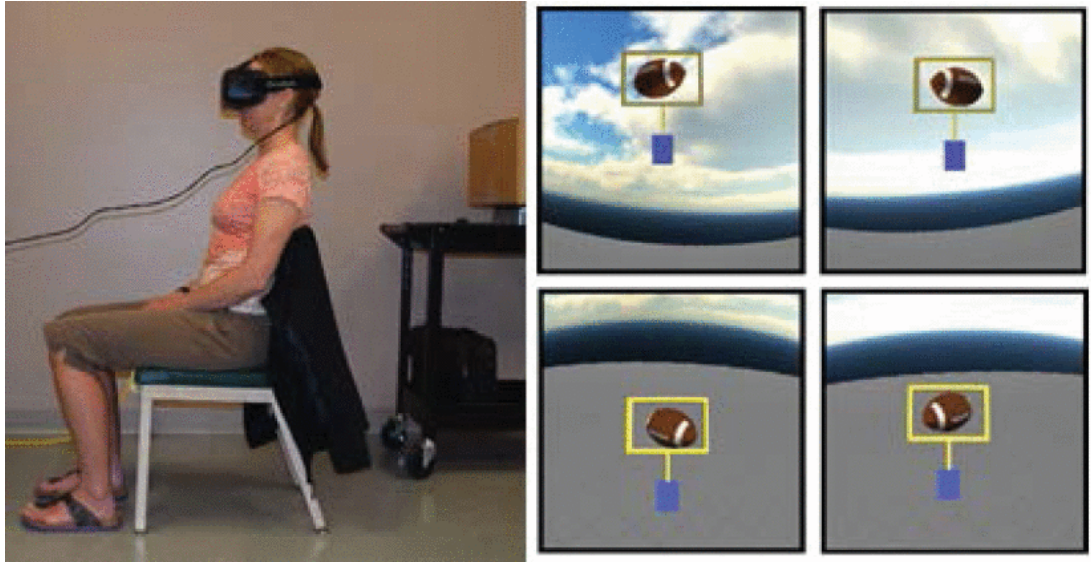

**Figure 12. Participant wore a 3D head-mounted display in front of a motion tracker (left). Four locations of the target image (right) [92]. Reprinted with permission.**

Figure 12 shows the experiment setup. The VR system comprised motion tracking camera and a 3D head-mounted display with inertial sensors for visual feedback and head orientation tracking. The four possible locations of the target football in Figure 12 were based on the average allowable range of motion in a diagonal head-neck movement. In the experiment, Chen et al. manipulated the visual feedback provided to individuals who performed physical exercise in VR. The altered visual feedback was intentionally designed to present a lesser degree of movement than the participant was actually moving. In this way, the participants actually moved more than they perceived or saw in the VR environment. Participants tended to tolerate the discrepancies to a certain extent between the visual feedback in VR and their own proprioceptive feedback. After analyzing the ratio between actual head rotation angles and visual rotation angle displayed in individuals with and without (asymptomatic) chronic neck pain, the authors concluded that the visual-proprioceptive mismatch led to an increase of neck rotation movement in VR environment. Therefore, the altered visual feedback in VR can encourage patients with chronic neck/musculoskeletal pain and those with kinesiophobia to perform more rehabilitation exercises than normal exercises without the use of VR.

## 2.1.3 Applications of virtual and augmented reality technology in upper and lower limb prostheses

As described previously, recent decades have seen VR and AR models being incorporated into the pre-fitting prosthesis training phase. The VR and AR technology has brought many benefits to users, such as providing users with dynamic visual feedback, enhancing user interactivity by attaching superimposed prosthesis models onto the residual limb, and also lowering financial and labor costs compared with traditional rehabilitation training. The following text will explore the specifics of various VR/AR applications in upper and lower limb treatment, including the value they add to traditional systems.

The Computer Assisted Rehabilitation Environment (CAREN) (Motek Medical, Amsterdam, Netherlands) at Walter Reed Army Medical Center is a VR-facilitated prosthetic training and functional assessment system aiming to assist users with substantial training and evaluation procedures for quick function restoration with the goal to regain independence for daily activities of living [65]. The system includes a motion capture system (Vicon Motion Systems, Oxford, UK) equipped with ten cameras that capture the movement of users' bodies in real-time, a hydraulic motion platform with six degrees of freedom (DoF), and an arc projection screen that displays real-time visual feedback (Figure 13). During training, the range of motion (RoM) and the movement symmetry of the user can be recorded for an evaluation on a user's motion effectiveness. Users can also adjust their body movement in real-time based on the visual feedback on the screen during the training process, which improves their symmetry and alleviate the workload on their sound arm [65].

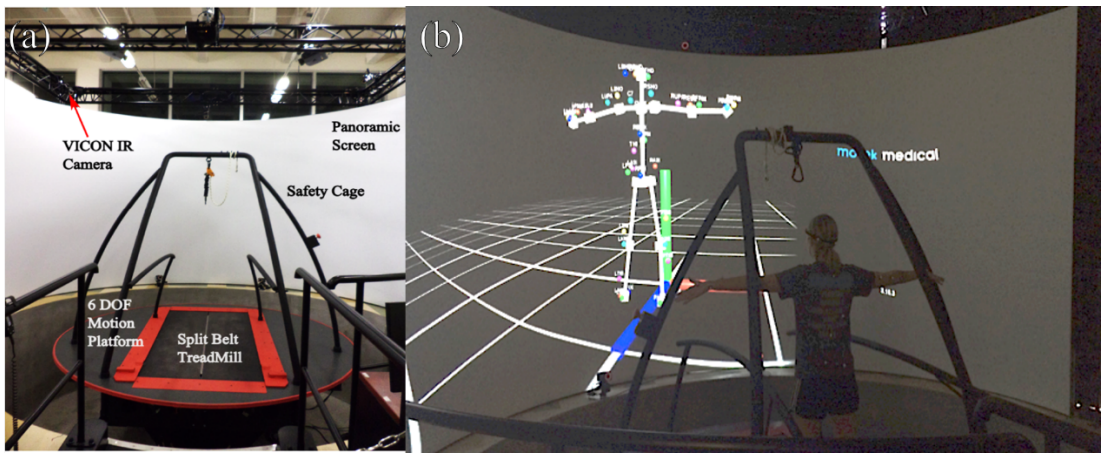

Figure 13. An overview of the CAREN system. Fig. 13 (a) shows the system composed of a 6 DoF motion platform together with a 180° cylindrical screen to provide an immersive virtual reality environment. Fig. 13 (b) shows he real-time interaction with CAREN system. Subjects perform tasks based on the real-time visual feedback from the virtual reality environment [65]. Reprinted with permission.

Research conducted by Markovic et al., resulted in the development of an AR-based interface for a closed-loop semi-autonomous prosthetic arm (Figure 14). The system is aimed at helping amputees to restore grasping functionality. AR glasses with stereovision cameras (Vuzix, Rochester, NY) scan the target object and process its geometry to determine the type of grip and aperture size required for the prosthetic hand to appropriately interact with it. The AR glasses then visually prompt the user so that they may adjust the posture and aperture of the hand before the system executes the grasping function. While posture and aperture adjustments are achieved through MPR, grasp type determination are determined autonomously with the visual information from the AR glasses. This pre-processing reduces the cognitive load on the user, providing more natural control of the artificial limb [93].

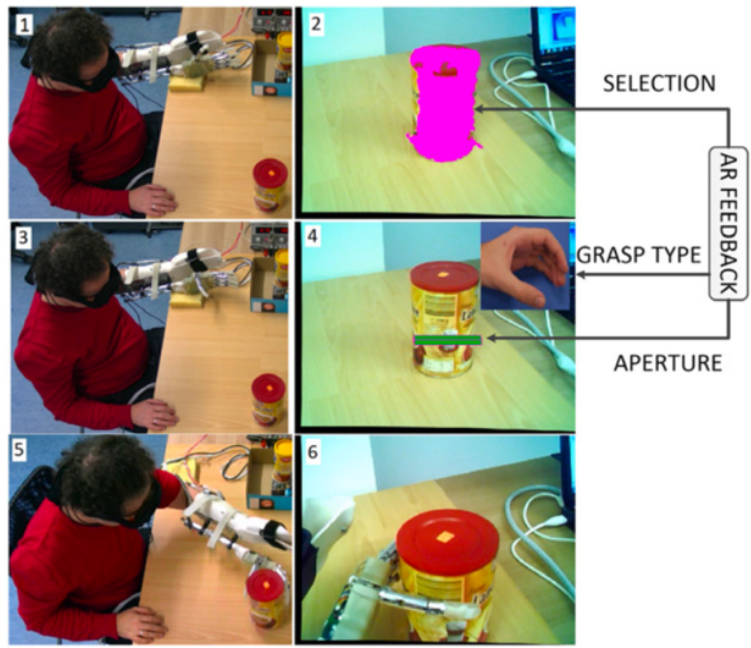

**Figure 14.** Semi-autonomous control of the prosthesis based on AR interface. In the beginning, stereovision camera on AR glasses will scan the size and shape of the target object and upload the information to computer vision-based algorithms for data processing. The algorithms will generate the output of upcoming grasp type and aperture size of the prosthesis and show the information to users on the AR glasses. Users can switch the grasp type and aperture size through sEMG control by performing flexion/extension muscle activities [92]. Reprinted with the permission of © 2014 IOP Publishing Ltd.

The paradigm of AR and VR has also been applied to propose and evaluate new functional outcome measures and to provide quantitative training tools aimed for pre- and post-fitting of a prosthesis. The tools currently available for functional assessment of patients using all levels of technology in upper limb prosthetic devices are severely limited. Current outcome measures are mostly qualitative in nature, which, although useful, are limited to a small workspace, do not accurately represent "real-world" utilization, and do not quantitatively measure the amount of body compensation during tasks.

To address such limitations, Sharma et al., established an AR-based functional outcome measure to assess upper limb amputees referred to as HoloPHAM (Holographic Prosthetic Hand Assessment Measure) [94]. Based on the physical PHAM platform (Figure 15), they are both comprised of an adjustable windowpane structure that houses a variety of geometric primitives. These geometric primitives may only be manipulated with specific hand grips and are meant to simulate activities of daily living. A stereotypical PHAM task to be executed by a prosthesis user consists of reaching to one of these objects, grasping it, changing its orientation, and relocating it within a target receptacle in a separate area of the user's reaching space [95]. During these tasks, user kinematics and dynamics are measured via multiple sensing instruments, including inertial sensors and a pressure mat. Post-hoc analysis of this information then allows a PHAM user to receive a score that quantifies the efficacy of their motion during assessment.

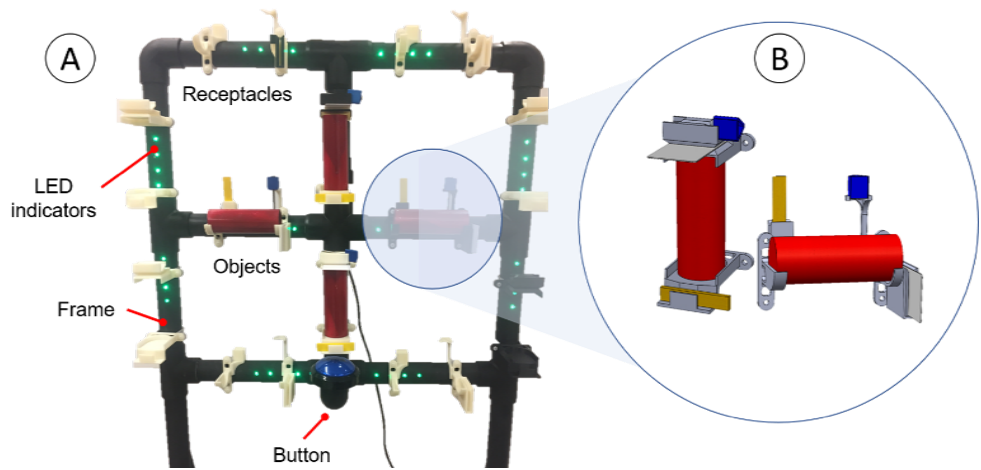

**Figure 15. A) Detail of the PHAM frame. B) Detail of the grip specific objects (red cylinder, blue prism, yellow block, and white card) and receptacles (gray). Note that the horizontal and vertical object receptacles are not the same. Distinctive designs are necessary to reduce the chance of collision while still guaranteeing each object can only be manipulated with one grip [95]. Reprinted with permission.**

As an extension of the PHAM, the HoloPHAM is aimed to help the amputee practice control of their myoelectric signals through structured object manipulations at home, as opposed to in a clinical environment. As with the actual myoelectric prosthesis, sEMG signals are fed into a MPR system which then decodes and classifies those signals into hand postures. These commands are sent to the AR interface to control the movements of a virtual prosthesis displayed in physical reality. A HoloLens MR system (Microsoft, Redmond, WA) is used to display the augmented environment, using the PHAM's network of inertial sensors to track the 3D motion of the residual limb to determine where to project the virtual prosthesis (Figure 16). The augmented scene consists of a first-person view with the virtual prosthetic limb projected on top of the residual limb, along with the various virtual objects that comprise the PHAM.

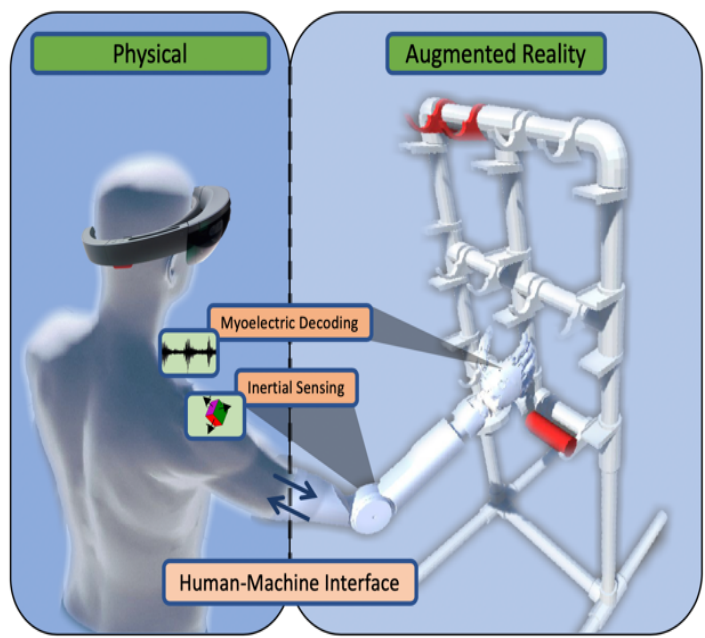

**Figure 16. The HoloPHAM system for upper limb myoelectric prosthesis control and training. Interface with the virtual prosthetic limb is through inertial sensing and myoelectric decoding that control the kinematics and grip patterns of the limb, respectively. The AR environment is projected through the Microsoft HoloLens, where the**

**subject receives both visual and tactile feedback during interaction with the virtual objects [94]. Reprinted with permission.**

The HoloPHAM system allows users to perform the same set of tasks as they could with the physical platform while also providing visual and haptic feedback to cue virtual object interactions. As with the physical platform, HoloPHAM also records user kinematics to be processed into quantitative motion metrics for evaluation.

Another approach for non-immersive VR-based amputee training is proposed by de Lima et. al [96]. The proposed system was initially designed to work with body-powered prosthesis users. Mechanical in nature, this kind of prosthesis generally consists of a system of tethers and harnesses to actuate the prosthetic hand (Figure 17a). In this study, the equipment consisted of a wire attached to a ring. In turn, the ring was aligned with the center of the user's back and attached to the contralateral shoulder with respect to the prosthesis. In this way, when moving the contralateral shoulder, the user applies a force on the cable that pulls the prosthesis' hand open. Their novelty relies on a sensorization that captures the user´s intention to open or close the prosthesis' hand (Figure 17b).

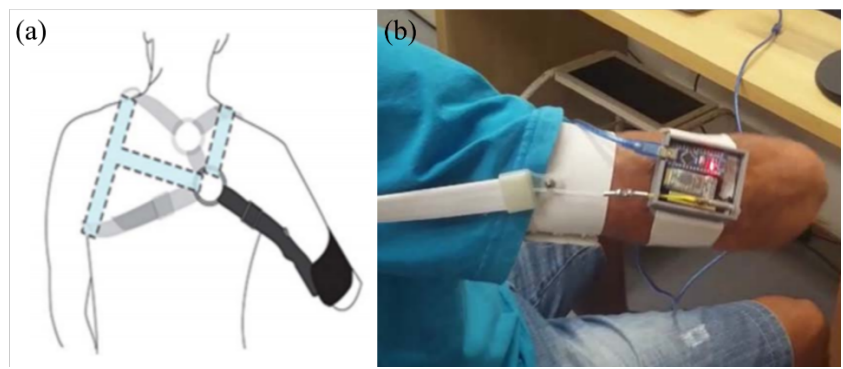

Figure 17. (a) Tether model. (b) An amputee using the equipment with sensors. With the movement of the other shoulder, the wire is displaced, resulting in the sensor's potentiometer sliding and a measurable voltage difference [96]. Reprinted with permission.

The user intent sensor consists of an Arduino Nano combined with a 10 DoF GY-88 inertial measurement unit containing two motion sensors: an MPU6050, a 6-axis gyroscope and accelerometer, and a HMC5883L, a 3-axis digital compass. Using an onboard sensor fusion algorithm, these sensors allow the device to identify its position and orientation in the real environment. With this information as feedback, it is possible to trace the position of the individual's residual limb in the real environment and send that data to the computer to determine the coordinates of the virtual prosthesis. Additionally, a linear sliding potentiometer is connected to the tether cable. Therefore, when the user moves the contralateral shoulder, applying a force on the tie rod, the potentiometer cursor slides in the same direction and the value of that displacement is sent to the application, which updates the virtual scene by opening or closing the virtual prosthesis in the same proportion.

To evaluate the system, the authors also developed a serious game where the user is prompted to collect objects of different sizes and shapes and put them in a box. To collect the objects, the users have to train how to open and close the virtual prosthesis. The system has demonstrated its potential to minimize the time to adapt and to control a real prosthesis. Minimizing the time to adaptation is crucial since most amputees take a long time to master a physical prosthesis, with some losing motivation and giving up during the training process.

Despite the contributions provided by this work, two major limitations arise. Firstly, the system is non-immersive. This reduces the naturalness needed for device embodiment. Secondly, the user has no feeling of owning the prosthesis. In fact, the virtual prosthesis is located in a different environment from that of the patient, resulting in a lack of intuitiveness. Facing these drawbacks, Lima et al. have recently proposed an immersive AR-based serious game for amputee training [96]. The serious game and the device to capture user intention are both based on the work proposed in [97]. The system also simulates a body-powered prosthesis. Since this approach uses a Microsoft Hololens for its visual feedback, this approach allows the placement of a virtual prosthesis on the amputee's actual residual limb, increasing their feeling of ownership. In order to track the amputee's residual limb movements, a circular marker was designed (see Figure 18). This work demonstrates that the integration of serious games with immersive AR presents as an effective paradigm for amputee training in the near future.

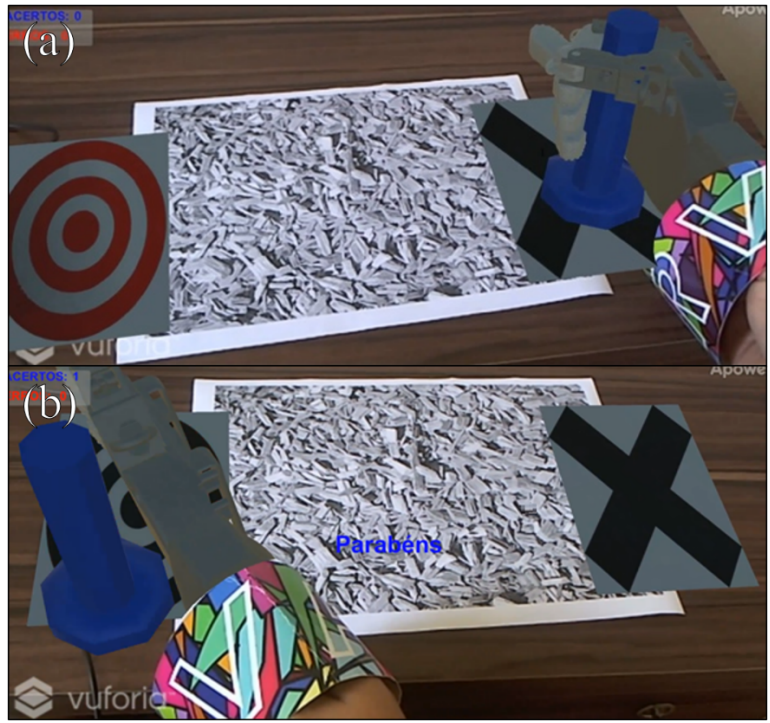

Figure 18. (a) User grabs a virtual object at the starting point. To gain points, they must take the blue object to the target point. In this scene, the table, the circular marker and user´s arm are real. Objects related to the game are virtual. (b) The user drops virtual object at the target point. Note that the user´s arm changes position (rotates), but the tracking remains, providing a faithful visual feedback [97]. Reprinted with the permission.

## 2.1.4 Applications of virtual and augmented reality in neuroprostheses

With the emergence of brain-machine interfaces (BMIs) in recent decades, hybrid neuroprosthetic devices show promising results in restoring functional mobilities for patients suffering from severe neuromotor deficits caused by stroke, spinal cord injury, cerebral palsy, and more [98]. By sampling and decoding the neural signals from the motor cortex through invasive or non-invasive BMIs, users can control neuroprosthetic devices with their intentions. The intervention of virtual and augmented reality technology into the field of neuroprostheses can assist users in getting accustomed to the use of neuroprosthetic devices and boost the motivation of users during practice.

Developed by the Johns Hopkins University Applied Physics Lab (JHU APL) as part of the Defense Advanced Research Projects Agency (DARPA) Revolutionizing Prosthetics, the Modular Prosthetic Limb (MPL) aims to facilitate prosthetic users with recovery of functions and abilities to perform ADLs [99]. The MPL system has 26 independently controllable movement classes (DOF) controlled by 17 independent motors, which meets most of the arm, hand and finger motor activities potentially required by users. The control of MPL can be achieved through the decoding of neural signal recordings sampled from the sensorimotor cortex of both hemispheres by microelectrode arrays (MEAs) [100]. Figure 19 shows the scenario of neuroprosthetic training of MPL in virtual reality. By intensively thinking about the control of the motion of MPL, those generated EEG signals can be decoded into specific commands to drive the motion of the virtual MPL on the screen, which enhances users' ability to control the neuroprosthesis.

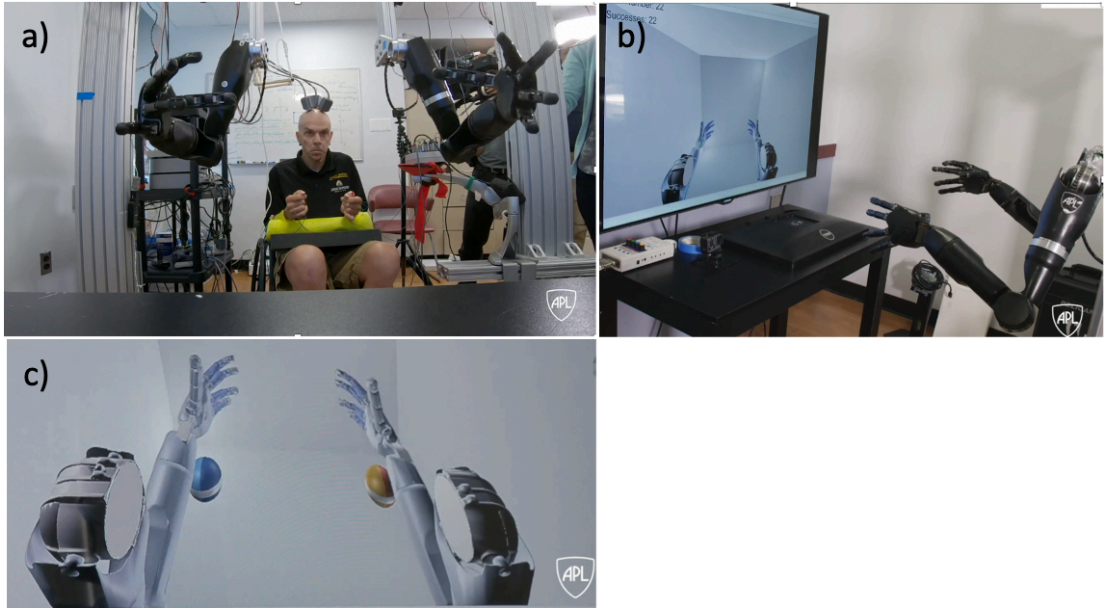

**Figure 19. MPL training in virtual environment. By capturing the neural signals through the implanted BMI device implanted on the participant's bilateral sensorimotor cortices and decoding it into specific control stream, the system can help patients perform the intended motor activities and help them with the practice of the neuroprosthesis control through the assistance of virtual reality [100][101]. Reprinted with permission.**

Another study that applies the technique of virtual and reality into neuroprostheses field is the Hybrid Augmented Reality Multimodal Operation Neural Integration Environment (HARMONIE) system developed by Katyal et al. [102]. The system integrates multiple technologies, such as computer vision, brain-machine interfaces, eye tracking and so on for a semi-autonomous control, which reduces the cognitive burdens of users facilitated with these assistive modalities (Figure 20).

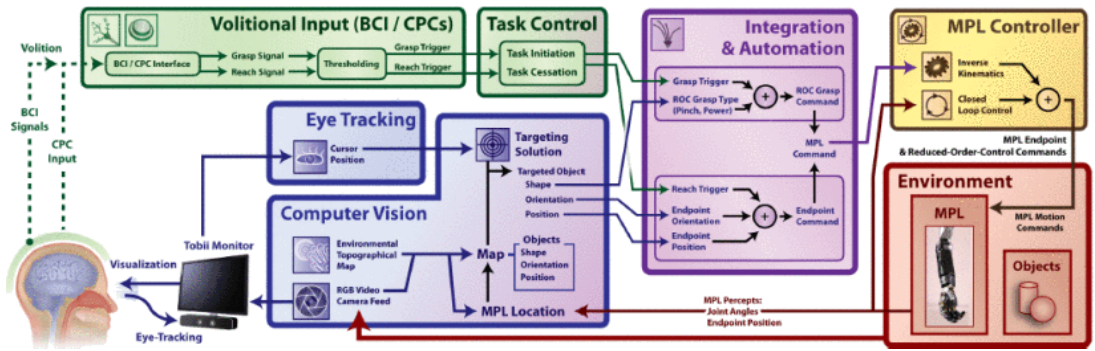

Figure 20. An overview of the HARMONIE system. The system applies multiple interface modalities, such as BCIs and eye tracking, for input collection and computer vision-related algorithms to assist users with prosthetic control intent integration and automation. Users can manipulate the prosthesis to perform tasks based on the visual feedback in augmented reality by Microsoft® Kinect [102]. Reprinted with permission.

The system uses Microsoft® Kinect (Redmond, WA) device to capture and identify the information of objects regarding their spatial position, shape and so on to assist with the decision making for grasp type and trajectory planning. And besides eye tracking modality they use for user interface control, the system also integrated other interfaces to assist users with prosthetic control, such as EEG signal sampling based on BCIs to interpret users' intentions towards the upcoming actions of the prosthesis, electrocorticography (ECoG) signal sampling for initiation of tasks corresponding to reaching activities, and sEMG signal sampling for acquiring physiological intent for controlling the prosthesis. And compared with the traditional joystick control, this hybrid approach is shown to outperform it with a 65% decrease on the total time consumed to perform grasping tasks, along with a smoother trajectory of the limb with the assistance of computer automation, which demonstrates that HARMONIE system can improve the performance with a shorter task completion time and less cognitive effort needed from users.

| Study | VR/AR paradigm | Application area | Number of subjects n | Age | Evaluation Metrics | Results |
|---|---|---|---|---|---|---|
| Cyrino et al. [84] | VR | Post-stroke treatment | - | - | Three-dimensional trajectory | The study shows the feasibility along with the advantages of the system, such as customizable level of challenge, inspiration for fun and engagement of users, trajectory of limb movements available for clinical analysis and evaluation of mobility |
| Assis et al. [85] | AR | Post-stroke treatment | 4 subjects with stroke | 49-61 yr | Ranges of motion, Fugl-Meyer scale | An increase in range of motion both in the AR training group (46.7% to 73.9%, +27.2%) and in the conventional group (61.3% to 90%, +28.7%); Statistically significant increase on the Fugl-Leyer scale for the AR training group (p<0.05), |
| AlMousa et al. [87] | VR | Post-stroke treatment | 5 patients with stroke | 56-61 yr | Time consumed for different level of challenges, number of errors or assistances in each task, subjective questionnaire | More time consumed in non-VR games, the statistics from subjective questionnaire showed positive trend towards VR system. Proved the feasibility of the VR assisted training system for post-stroke treatment, along with the level of engagement and motivation it provides to users, |
| Chen et al. [92] | VR | Chronic neck pain and kinesiophobia | 9 participants with chronic neck pain | 39-75 yr | Self-reported pain level, Tampa Scale for Kinesiophobia score, ranges of motion, control-display gain (ratio between actual head rotation angle and visual rotation angle displayed) | Self reported pain level of 2.1± 1.3 on the VAS (0 =no pain, 10 =maximum pain). The Neck Disability Index was 15.3± 4.7 (0 =no disability, 50 =maximum disability). The mean Tampa Scale for Kinesiophobia score was 23± 5.9 (0 =not kinesiophobic, 44 =very kinesiophobic). The mean left and right axial rotation ROMs were 61.8◦ and 63.4◦, respectively, demonstrated the feasibility of |

| | | | | | | virtual reality system in chronic neck pain treatment |
|---|---|---|---|---|---|---|
| Isaacson et al. [65] | VR | Lower limb rehabilitation | - | - | - | Demonstrated the advantages of CAREN system for lower extremity treatment, such as being customizable for task arrangement, physiological measurement available |
| Markovic et al. [93] | AR | Upper limb rehabilitation | 13 able-bodied subjects | 25-33 yr | Task completion success rate, task completion time, grasp control failure rate | Statistically significant decrease in the time consumed for completing the task and increase in the task completion success rate (81%) in Semi-AR system, which showed the effectiveness of the semi-autonomous prosthetic control system |
| Sharma et al. [94] | AR | Upper limb rehabilitation | 3 able-bodied subjects | - | Task completion time, | Statistically significant improvement in the reduction of task completion time in HoloPHAM (28.5%) compared to PHAM(23.5%), which showed the effectiveness of the AR-assisted system |
| Cavalcante et al. [97] | VR | Upper limb rehabilitation | 7 able-bodied subjects, 1 amputee | - | Post-task questionnaire | Data from the questionnaire validated the usability, functionality, efficiency of the system |
| Yu et al.[99] | VR | Upper limb rehabilitation | - | - | - | Demonstrated the benefits and functional advancement of the neuroprosthesis system |
| Katyal et al. [102] | AR | Upper limb rehabilitation | 1 subject | - | Time consumed to reach and grasp objects in the task, trajectory of the prosthesis | The average task completion time by using the hybrid system decreases by 65% compared to traditional joy stick control. The trajectory under this approach is also the most smooth and consistent indicating the minimal distance cost for motion. |

**Table 1. An overview of the current VR/AR applications in neurorehabilitation. The table below systematically summarizes all the studies and applications covered in this section, including the paradigm they apply, the target issue they address, number and age of**

**subjects participated in the study, the metrics used for assessing the applications, along with the results of the studies.**

## 2.2 Discussion

This section first holds a discussion about the pros and cons of VR/AR assistive tools applied in neurorehabilitation field, and explores some of the wider-ranging, practical benefits of augmented and virtual reality technologies on neurorehabilitation not yet mentioned. Additionally, this section outlines some of the technology's current limitations and avenues for future improvement.

### 2.2.1 Overview of VR/AR assistive tools for neurorehabilitation

In the previous section, we demonstrated the use of VR/AR technologies in neurorehabilitation field by introducing several representative examples of VR/AR applications in various subfields, such as post-stroke treatment, kinesiophobia, upper- and lower-limb rehabilitation and neuroprotheses fitting. In this section, we will delve further into the critical analysis of VR/AR techniques in this area by highlighting the pros and cons of some cases selected from the above based on the results or informative data from the studies.

One of the advantages that VR/AR-based applications boost in neurorehabilitation area is the level of engagement and entertainment they bring to users during treatment, which, as mentioned before, is a crucial aspect for users to persist and accomplish their treatment plans. Giving an overview of these cases from a comprehensive perspective, we find that most VR or AR assistive tools served for rehabilitation purposes can entertain users in format of games or interactive motor activities, which will help motivate participants to stick on to their training program and achieve better efficacy in rehabilitation. Aided for post-stroke rehabilitation, the training interface HarpyGame developed by Cyrino et al. can successfully bring entertainment to users while performing motor tasks by correlating them with the control of the avatar eagle in the VR game to achieve specific goals such as avoiding virtual obstacles, flying through target routine and heading towards food [84]. Similarly, the Move-IT system developed by AlMousa et al. also get users practice their mobility tasks for post-stroke rehabilitation by interactively playing the colored cube-sorting game [87]. And according to results from the post-task subjective questionnaire, 4 out of 5 participants agree with the opinion that it is comfortable and encouraging to play the game for rehabilitation exercises, and all of them indicate that such paradigm would support their rehabilitation [87]. One technical feature to notice is that compared to HarpyGame as a non-immersive VR interface which uses a monitor or screen to cast the displayed content, Move-IT can provide trainees with an immersive experience through the head-mounted display (Oculus Rift, Facebook Technologies, LLC, USA). Based on the three-dimensional visual feedback from the headset, users tend to be more engaged in the tasks compared to non-immersive methods. However, such immersive visual display also exists limitations that potentially draw negative impacts on user experiences, which may bring concerns to certain populations. Some people not accustomed to the dynamic visual feedback from immersive display may feel dizzy or sick when exposed to the display. Therefore, it would be arbitrary to conclude one over another when comparing the two display formats, and different aspects, such as human factors and the target area, should be taken into consideration when choosing a display format from them for different applications.

Another remarkable point for these VR and AR applications for rehabilitation purposes is that the content can be customized or adjusted based on clinical needs or expectations. Such features can be specific in adjustable level of difficulty or challenges to provide users with a self-paced rehabilitation experience, or the flexibility in simulating a designated scenario for trainees to practice tasks that are hard to accomplish due to various factors such as high cost, potential risk and so on. From the examples mentioned earlier, VR gaming systems such as HarpyGame and Move-IT for post-stroke treatment and the upper-limb training system by Lima et al. allow users to switch to other types of tasks or change the level of challenges in the training program, which will fit the needs of people with different needs and situations properly [84][87][96]. In addition, the CAREN system for lower-extremity treatment can also provide users with customizable tasks that allow users to get trained in virtual scenes that simulate daily-life environments. Instead of having to getting exposed to roads or streets that are risky for trainees to practice their gait and may bring injuries, they can get themselves trained in a simulated environment safely and effectively [65]. Meanwhile, the CAREN system also offers different formats of immersive visual display ranging from two-dimensional flat video to a 360-degree surrounding scene, from which users can choose the specific display format based on their needs. However, because of the naturalness of the scale of the scene that simulates daily-life situations, a large enough space for the training may be required for immersive VR display, which may potentially raise concerns to general populations in need for the training.

Furthermore, quantitative assessment and evaluation for users' performances in doing motor tasks can be conducted accurately and with the facilitation of different techniques such as motion tracking, eye tracking, computer vision and so on, more informative data that reveals users' activities from different perspectives can be available to users, which can be helpful to assist therapists on making clinical decisions or adjusting treatment plans for patients. Some preliminary examples include the VR gaming systems such as Move-IT system for post-stroke treatment, where participants' performance in doing motor tasks are conveyed through the time spent in different levels of difficulty of the game, the total amount of failures, error rate and call for assistance during the game [87]. Other cases such as HarpyGame can track the trajectory of the avatar eagle movement in virtual environment and by comparing it with the designated route in the game, therapists can get to know the performance of participants in controlling their arm for motor tasks (Figure 21) [84].

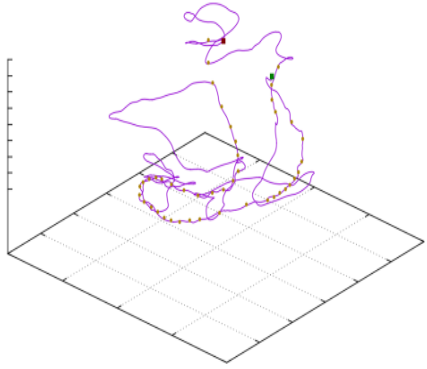

**Figure 21. Harpy eagle's trajectory graph generated from one patient session. The green point indicates the start of the training whereas the red one indicates the end. Other points show the challenges along the way [83]. Reprinted with permission.**

Benefitted from the incorporation of motion tracking technology, the CAREN system and HoloPHAM can capture the movement of participants' body parts through those integrated motion tracking devices such as data gloves and inertial measurement unit (IMU), and these kinematic information can then be interpreted into quantitative metrics that are reflective when assessing users' performances, such as 3D trajectory of joint dynamics during mobility tasks implementation, efficiency of the movement distance compared to optimal path and so forth [65][94]. Moreover, in order to obtain a comprehensive assessment from various perspectives to provide clinical analysis with more references, many more paradigms can be combined into the assessment systems for acquiring and processing input data from different aspects. Besides what has already been mentioned in CAREN system regarding its assessment metric about mobility, many physiological features can also be quantitatively measured such as heart rate, VO2 and other ventilation data [65]. By utilizing different wearable sensors and integrating them into the VR system for data sampling, more physiological information regarding participants' biophysical state during their training can be revealed and made available for analysts to look into. Yet these contents are discussed in the range of feed-forward layer for a system where emphasis are laid on getting inputs from subjects. In fact, feedback also plays a significant role in assisting users with rehabilitation. By transmitting the corresponding feedback during the rehabilitation treatment, users can adjust the motor activities of the body to fit themselves to the training properly. Apart from the visual feedback provided by VR or AR interface, other feedbacks that are conveyed to different sensing area of human bodies can not only provide more intuitive assistance to users, but also helping them better accustomed to the simulated virtual environment as well, which enhances the level of immersion of the system and alleviates the ambiguity compared to those only provides visual feedback. The HoloPHAM system can both provide haptic feedback to subjects during their usage. When touching a virtual object with their virtual prosthetic hand, vibrotactile sensors will vibrate and generate haptic feedback for the subject's arm to sense, which simulates the feeling in the real-world when the subject interacts with an object with a prosthetic arm.

### 2.2.2 Financial and labor costs in neurorehabilitation

Neurorehabilitation therapy typically requires a significant amount of financial and labor investment. Patients will go through the rehabilitation process by working together with other healthcare staff and rehabilitation professionals who provide guidance and clinical care from various perspectives [71]. Such an interdisciplinary team may involve the participation of physiatrists, surgeons, and nurses (as professionals who provide intra-surgery and post-surgery healthcare from the clinical aspect), along with certified prosthetists, behavioral health specialists, physical and occupational therapists (OTs), recreation therapists, vocational rehabilitation specialists, driver rehabilitation therapists, and so forth [34]. Among the professional staff from the rehabilitation center, OTs are the ones that are highly involved and devoted to promoting and supervising the neuromotor training regimen. An individual with a neuromotor deficit is highly dependent on the guidance and instructions given from OTs and/or prosthetists to learn how to generate voluntary muscle activity to accomplish tasks in daily living [72][73]. Therefore, a large amount of human labor is involved in the training process. This process also demands a lot of effort from therapists and prosthetists to teach end-users, especially individuals using MPR control systems. The financial costs to involve these professionals are usually high, yet they are not always available when trainees need guidance and instruction [74]. According to a survey investigating upper-extremity and lower-extremity amputees' satisfactory level towards rehabilitation services, only 25% of them claim to have gone through rehabilitation services in the past year. Around 20% of them indicate that they are often unable to access help and care from professionals when needed, with financial limitations considered the most common limiting factor [75].The intervention of virtual and augmented reality interfaces can assist users with neuromotor deficits throughout the training procedures by guiding users to follow the system's instructions to perform motor tasks. Such technology may dramatically improve the therapy's outcome while optimizing the therapists' work in supervising and instructing trainees [76]. What is more, the financial cost of game-based training tools with the support of VR and AR are also found to be lower in comparison with the traditional rehabilitation methods [77][78].

### 2.2.3 Portability and accessibility of neurorehabilitation

The availability of a portable neurorehabilitation paradigm can positively impact the quality of therapeutic services. The majority of current protocols related to conventional neuromotor training lack portability and are restricted to rehabilitation centers. Participants need to routinely head back and forth between their daily-living locations and the rehabilitation institutions, which indirectly effects the completeness and efficacy of the rehabilitation training regimen. According to the survey conducted by McCarthy et al., about one-third of the amputee patients who suffered from upper-limb amputation do not reach the appropriate number of visits to the therapy center [79]. The amputees' lost motor functionality limits the frequency of their visits to a rehabilitation center. Coupled with the fact that effective training requires nearly daily therapeutic intervention, participants often experience lackluster progress during their pre-fitting training.

Typically consisting of a display device, a personal computer, and other human-computer interface devices such as myoelectric control chips, electroencephalographic cap kit, or simple motion trackers, a VR/AR-based neuromotor training system is much more portable and allows users to train at home or work, at any preferred time without restriction. Such self-paced training offers not only personal convenience and flexibility to users but also substantial training efficiency while reducing the user's reliance on healthcare professionals.

# Chapter 3. A comparison between virtual reality and augmented reality on upper-limb prosthesis control

In this chapter, we conducted a study focusing on a comparison between virtual and augmented reality system regarding their effects on upper-limb prosthesis control. In this study, we aim to determine which paradigm, AR or VR, is better suited for the completion of dexterous motor control tasks needed for effective upper-limb prosthesis use. This work has been presented in the 2021 International Conference on Frontiers in Digital Signal Processing and is undergoing the publication process by ACM Digital Liberary.

## 3.1 Introduction

Limb loss addresses much inconvenience to victims and therefore a serious issue to consider. According to the statistics published by the National Limb Loss Resource Center, it is estimated that there are around two million individuals suffering from limb loss each year in the United States, yet the number is estimated to double by the year 2050 [13][14]. Among the people suffering from limb loss, nearly 185,000 individuals receive amputation as treatment for a medical condition each year. Upper-limb amputation is considered to be extremely challenging in particular. Upper-limb amputees encounter not only functional limitations, such as disability to perform grasping to handle basic tasks in daily life, which negatively influence their individual independence, but also from a psychosocial perspective as well, such as the loss of capability to perform communicative gestures and sensing [15][16]. Besides the functional inconvenience brought to the living situation by the loss of limbs, there are potential sequalae after amputation surgery that might bring a sense of discomfort to amputees both physically and mentally, such as phantom pain [107].

One of the methods to assist amputees with handling these issues and alleviating those negative influences is to equip them with a prosthesis, among which myoelectric prothesis is a technically advanced option. Basically, the motor control of the prosthesis is achieved by the different kinds of contractions generated by the residual muscle group. Surface electrodes from the prosthetic device attach to the certain residual muscle group of the user to sample the surface electromyographic (sEMG) signals, which are then interpreted through proper methodologies, such as pattern recognition, into specific classifications on different grasping logics that drive the motivational activities of the prosthesis as a consequence [108]. However, there's always an ineluctable time period lying between the time an amputation surgery is operated and the time amputees receive their customized prosthetic limb ready for them to utilize. Additionally, in order to gain better control towards the different functional capabilities of the prosthesis, amputees would need substantial muscular training during which they need to learn how to contract different muscle groups distinguished enough in certain ways to coordinate certain prosthesis functions, such as hand open/close.

Traditional rehabilitative prosthesis training methods involve therapists in a rehabilitation center instructing trainees to voluntarily activate their target muscle sites in preparation for the future use of a myoelectric prosthesis. This is monotonous and yet not intuitive, as trainees cannot get any feedback or interactives with a prosthetic arm from the training process and thus may result in difficulties in manipulating a prosthesis when they are in usage of the device [22][23][24].

To tackle such issues in rehabilitative training, many prosthetic device developers and investors have been implementing innovative prosthesis training methodologies. Both hardware and software technologies have been incorporated to improve the myoelectric prosthesis training efficiency and strengthen trainees' learning experience. An example is the Myoboy® software suite carried out by Ottobock

Company (Ottobock HealthCare, Duderstadt, Germany) in 2011 to guide users in prosthesis training in the format of playing a simple game with biofeedback [10]. In recent years, with the trend of virtual reality (VR) and augmented reality (AR) being popular and widely applied as immersive technologies, prosthesis researchers and developers have begun to apply such technologies into prosthesis training that could provide amputees with an immersive prosthesis training environment virtually, which will consequently help improve users' training experience and reduce the burdens of both amputees and clinical parties [38]. Users are able to get their control ability well trained with proper visual feedback and without suffering from the weight of the prosthesis, whose physical version is relatively heavy and may cause discomfort to amputees [39].

However, though the rehabilitative prosthesis training in virtual environment, either VR or AR, may take advantages in portability for carrying along, convenience and efficacy for a self-paced training, uncertainties and possible improvements still lie in such approaches [22]. Furthermore, based on the essential differences between the conception and applied avenue of VR and AR, the impact and performance that the two different virtual environments generate in facilitating prosthesis control is in need to study and compare. In that way, occupational therapists (OTs) or trainees themselves would get a clear conception of the pros and cons of prosthesis training in VR and AR from various aspects, and thus reasonably choose the virtual environment that could be more helpful and fitful towards their rehabilitation training according to their personal circumstances and needs. However, to the best of our knowledge, seldom comparative studies between the performance and impact of VR and AR in facilitating prosthesis control have been carried out.

On the other hand, many exploratory research and directional studies have been carried out to compare the performance of VR with AR on their applications towards other medical involved areas, such as their applications in laparoscopic surgery, strategy-based mental therapies and so forth. For example, there is a paper focusing on a comparison between AR and VR on their applications towards exposure-based therapy for certain phobias [109]. The study conducted experiments and measurements on different dimensions to evaluate the performance of the different kinds of therapies. They came out with conclusions suggesting that AR-based exposure therapy outperforms VR-based version in certain aspects, such as higher spatial presence, that leads to better and quicker adaptation to the semi-synthetic-based therapy. But the measurement of the principal factors and the resource of the data relies relatively heavy on the subjective feelings of participants, which lead to a more empirical-based result and might contribute to subtle errors when applied in the evaluation of prosthesis training in virtual environments.

Based on the situations stated above, this paper will conduct a study focusing on the comparison between rehabilitative performance and efficacy of AR and VR-based virtual prosthesis control. We applied the functional outcome measure--Prosthesis Hand Assessment Measurement (PHAM) as the training prototype and built the virtual version of the tool in both VR and AR [95]. In the experiment, we evaluate a population of five able-bodied subjects, each performing a 3-dimensional object manipulation and relocation task in analogous AR and VR environments. Moreover, we applied kinematics-based quantitative measurement and analysis into the experiment, such as Fitts' law to evaluate the kinematic performance of the virtual prosthetic hand in a motor task along with path efficiency and task completion statistics.

## 3.2 Methods

### 3.2.1 Training Environment Requirements

To evaluate the feasibility of both VR and AR environments on prosthetic control, it is important to ensure that both paradigms provide the users with the same scenario facilities. Otherwise, the differences observed in this comparative study cannot be attributed solely to differences in the two paradigms. As such, the following requirements have been identified to ensure inter-paradigm comparability:

• Sense of Personification: the feeling of oneness with one's own virtual body [110]. It is the sense of being self-located. The user must feel as part of the environment. It is important that the user sees his own body and is able to control it. In virtual reality, the real body is dissociated from the virtual representation, but it has been shown that the sense of personification can be effectively achieved with the use of avatars [111]. In contrast to VR, augmented reality usually allows the user to see his real body and virtual objects are placed into his real environment, allowing direct interaction with these virtual objects. The sense of personification for this kind of application is not only important for the user experience of a virtual prosthesis, but it could also give the benefit of increased physical task performance, when the virtual prosthesis interacts with the virtual objects [112].

• Level of Realism: if the goal is to transfer task proficiency to the physical world after some time, the virtual or augmented environment must be similar to the intended physical environment. Therefore, the level of realism corresponds to the degree of convergence between the expectations of the user and the actual experience in the virtual (or augmented) environment [111].

• Immersion: refers to how accurate a given computer system is in providing the user with the illusion of a reality different from that in which he finds himself. That is, it is the objective level at which a virtual system sends stimuli to sensory receptors of the user [111]. In contrast to a VR user, an AR user does not leave the space he occupies. Instead, he is put in co-presence with virtual elements, blended into the non-synthetic world, giving the feeling of a parallel universe, out of his reality [109].

• Presence: refers to the feeling of actually being within the virtual environment and that the virtual environment seems to acknowledge the user by reacting to his actions [113]. Presence is a state of consciousness, a psychological perception of being in the virtual environment.

• Feeling of ownership: considering a virtual or augmented environment for upper limb training, it is important that the amputee feel the virtual prosthesis as part of his own body [111]. For VR environments, normally, an avatar with a virtual prosthesis coupled with a virtual stump is used. On the other hand, for an AR environment, the user has to see a virtual prosthesis coupled to his real stump. In both cases, ways to control the virtual prosthesis must be provided.

### 3.2.2 System Design

In order to evaluate and compare the efficacy of prosthesis control in virtual and augmented reality environments, the consistency of hardware devices applied in both scenarios during prosthetic training needs to be guaranteed. Therefore, in this experiment, we adopted the same set of hardware and software modalities to implement both environments. An overview of the system architecture is displayed in Figure 22. We used HTC Vive® Pro VR headset (HTC Corp., Xindian, Taiwan) for visual exhibition, along with two Vive® trackers (HTC Corp., Xindian, Taiwan) to track the motion of subjects' forearm. For the feedforward control, we used a Myo®band (Thalmic Labs, Ontario, Canada) to sample the sEMG signals of the residual limb generated during voluntary muscle contractions. We applied linear discriminant analysis (LDA) [20] to classify and decode the sEMG signals. In order to simulate the weight of wearing a prosthesis, we used a bypass prosthesis along with a 0.5 kg scientific weight attached to it. The software environment: Python 3.7.4 (Python Software Foundation, Dover, DE), and Unity3D® 2018.4.17 (Unity Technologies, San Francisco, CA) was installed and setup on the desktop for carrying out this experiment. For the hardware part, we used a computer with i7-10750H CPU@2.60GHz, 16GB RAM, and NVIDIA GeForce® RTX 2060 graphics along with Windows® 10 operating system to run the software system.

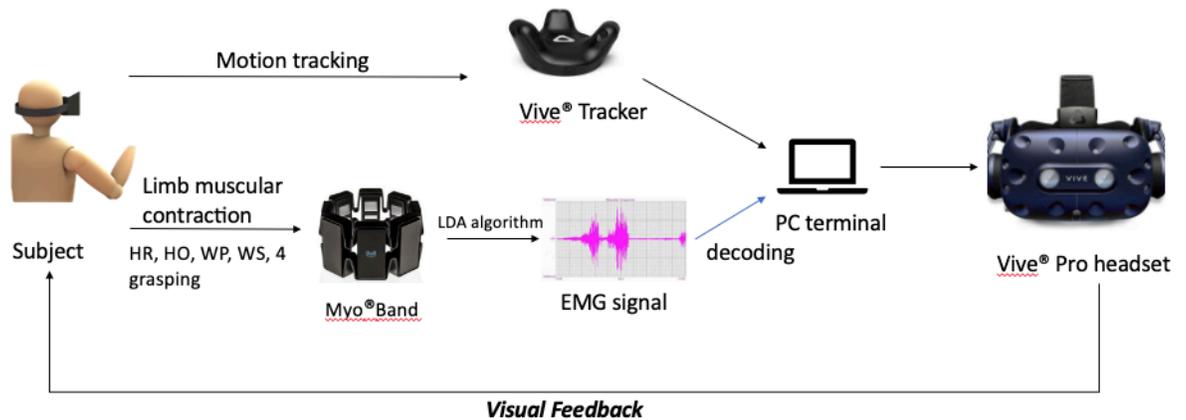

Figure 22. An overview of VRPHAM/ARPHAM workflow. The functional outcome measurement--PHAM is virtualized in both a VR and an AR environment. A Vive® tracker is attached to the right arm of a subject to track the position and the motion of the subject's arm in real-time; A Myo®band is placed next to the Vive® tracker to collect the sEMG data from the muscles and transmitted to the computer, which is then decoded into control commands to activate different grasps of the virtual prosthesis. The virtual prosthesis along with the PHAM frame will then be dynamically simulated in AR or VR and displayed in a Vive® Pro headset as visual feedback to the subject as a consequence.

## 3.2.3 Experiment Protocol

For this experiment, we recruited 5 able-bodied subjects, with an average age of 24 ± 1.5yrs to participate in the experiment. All subjects were right-handed and all manipulation tasks in the experiment were achieved by subjects through their dominant hand. The experiment was conducted with their informed consent. Before going through the experiment, subjects acknowledged the overall experiment protocol. The experiment protocol was approved by the Johns Hopkins Medicine Institutional Review Board. This protocol includes the experimental procedures that each subject goes through, along with the metrics used for evaluating the kinematic performances of prosthesis control.

### 3.2.3.1 Pattern Recognition Training

When subjects were ready for the experiment, they were first asked to don the Myo®band on their right forearm, at the position with maximum muscle mass. After placing the device firmly, they would spend around 5 minutes waiting for the electrodes to settle. Then subjects were asked to don the bypass prosthesis and affix a 0.5 kg scientific weight to the distal end to simulate the load of a typical transradial prosthesis. Once loaded, subjects were prompted through a standard pattern recognition training protocol, collecting data to train the LDA algorithm. Subjects performed a series of muscle contractions based on various image cues on performing different motor activities (hand open, hand rest, power grasp, key grasp, tripod grasp, pinch grasp, wrist pronation and wrist supination) prompted in succession on the computer screen. Each cue was prompted for 5 seconds, with the last 3 seconds worth of sEMG data being collected. The successions of cues were completed 3 times in 3 different arm positions: (1) at the side with a 90° elbow bend; (2) outstretched in front of the body at a comfortable posture; (3) outstretched in front of the body and raised above the head. If the cross-validated accuracy for any of the trained patterns were less than 80%, then the training phase would be restarted to ensure the stability of the myoelectric control. Every time when users took off the Myo®band and donned it again because of muscle fatigue, they would go through the pattern recognition training process again to ensure stable myoelectric control.

### 3.2.3.2 Pre-fitting Preparation

After going through the pattern recognition training phase, subjects were outfitted with the hardware necessary for VR and AR environment display and integration. It was verified by experiment staff that the two Vive® base stations covered the area where all the motion tracking-related devices (Vive® Pro headset, two Vive® trackers) might potentially reach during the experiment to avoid the potential circumstances in which any one of the devices might lose track. A Vive® tracker was placed on the ground to orient the virtual PHAM in the workspace while another tracker was placed on the subject's right forearm, to orient the prosthetic limb. The subject also wore the Vive® Pro headset to display the virtual environments, which is shown in Fig. 23. In order to mitigate the impact of learning potentially generated from the process of performing similar motor tasks in the assessment phase, subjects familiarized themselves with the experiment environment by performing a series of pre-fitting motor tasks that resemble the upcoming motor tasks in the assessment phase.

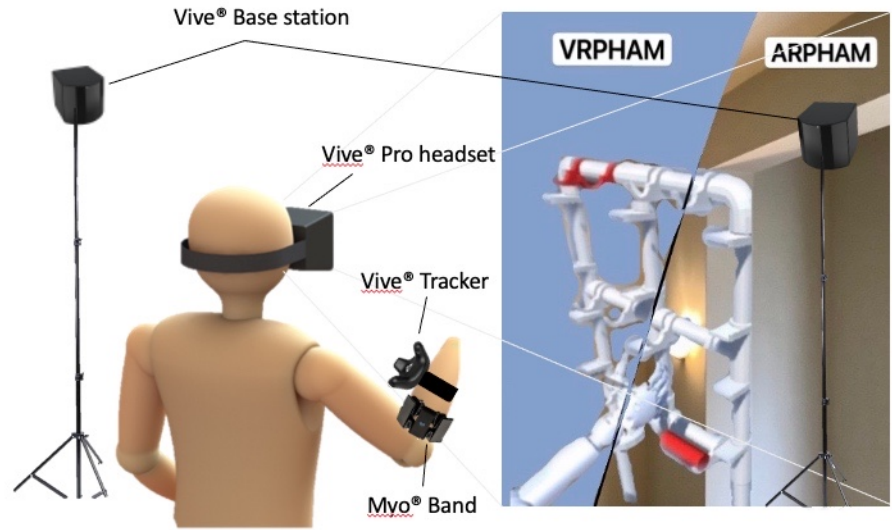

Figure 23. An overview of the experimental modalities setup. After donning the hardware modalities properly, the Myo®band and the Vive® tracker on their forearm and the Vive® Pro Headset onto their head, subjects will get familiarized with the AR or VR scene through pre-fitting tasks and then start the assessment phase of the experiment for data collection.

**3.2.3.3 Assessment Phase**

After going through the pre-training phase and allowing enough rest following the end of the pre- fitting phase, when subjects were ready, they would be allowed to go through the assessment phase. In the assessment phase, subjects went through two different groups of motor tasks: one in virtual reality and the other one in augmented reality. The order of whether VR or AR tasks were performed first randomized from subject to subject. Each group consisted of 5 consecutive sessions with 10 sequential trials as a session each time to reach and relocate the geometric primitives as cued by the system in VR and AR. After a session is finished, subjects were given 10 minutes to relax their muscles. After finishing assessment of in one paradigm, subjects would redo the pattern recognition training phase, as these control signals drift over time. For a single trial of object manipulation, a geometric primitive and a target location on the PHAM frame will be displayed in the virtual scene in VR/AR. Each subject would complete the following (Fig. 24):

1) Touch the button of the PHAM frame to signal the start of the manipulation task.

2) Control the virtual prosthesis to reach, grasp, and relocate the prompted geometric primitive to the destination holder.

3) Touch the button of the PHAM frame once again to signal the end of the manipulation task. This signals the end of a successful task.

If the object manipulation cannot be completed within 30 sec, the task will immediately terminate, and the next manipulation task will be loaded. This signals the end of a failed task.

Subjects repeated these steps for a total of 100 object manipulation tasks (50 in VR, 50 in AR).

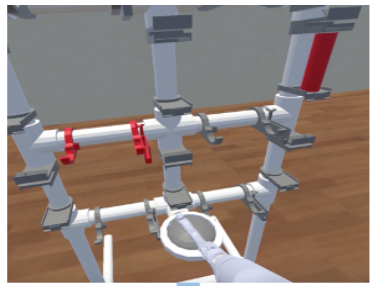
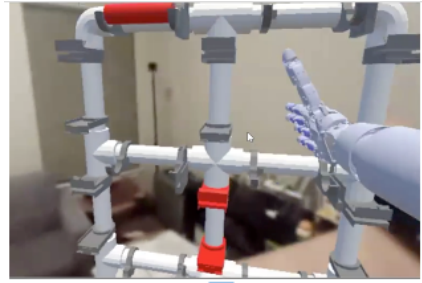

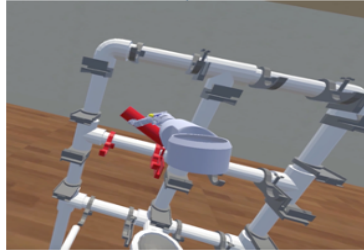
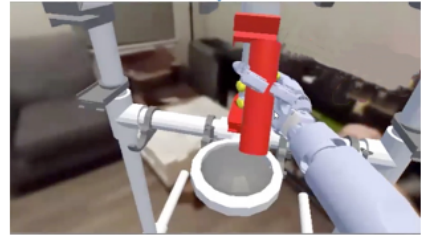

(a) VRPHAM　　　　　　　　　　　(b) ARPHAM

**Figure 24. Motor task manipulations in VR and AR environment. The figures above indicate the scenario of the motor tasks on object manipulations in VR and AR environment respectively. For a single object manipulation task, the following occurs: (a) A virtual geometric primitive out of four (cylinder, card, stick, triangular prism) is prompted on the frame of the virtual PHAM along with the target holder via visual cues at a random location. (b) The subject must reach and grasp the prompted geometric primitive through the virtual prosthetic hand. (c) Once grasped, the subject must rotate the virtual wrist vertically and relocate the object to the prompted target holder. (d) Finally, the subject must release the object at the prompted destination in the proper orientation. Each manipulation has a maximum allowed time of 30s before failure and subjects are asked to complete 50 manipulations in both AR and VR.**

**3.2.3.4 Segmentation of Phases**

In order to more granularly inspect performance of prosthesis control in VR and AR, each object manipulation was segmented into two phases: reach and relocation. Pressing the start button signals the beginning of the reaching phase and when the center of the prosthetic hand reaches within the bounding sphere of the target geometric primitive with a radius of 1cm, the reaching phase ends, and the relocation phase, lasting till the end of the task.

**3.2.3.5 Metrics**

In order to quantitatively measure and evaluate a subject's kinematic performance when performing motor tasks in VR and AR, we applied several evaluation metrics relevant to the task completion rate and time consumed in each phase along with kinematic efficiency in a single trial to quantitatively measure and evaluate the kinematic performance of subjects on task completion in a variety of aspects. First, we computed task completion rate (CR), which is computed as the ratio of the number of successful tasks (Nsuccess) and the number of total tasks (Ntotal), which is:

$$CR = \frac{N_{success}}{N_{total}} \tag{1}$$

Additionally, we measured the movement time (MT) consumed in reach and relocation phase in a successful trial. For the kinematic analysis, we applied Fitts' law to quantify the level of difficulty of each task along with the index of performance to assess the manipulation efficiency of a task [114]. For each successful task, the index of difficulty (ID) quantitatively reflects the level of difficulty of a target movement task which can be calculated as:

$$ID = \log_2(\frac{D}{W} + 1) \tag{2}$$

where D is the L_2-distance from the initial location of the prosthetic hand to the target location, and W is the tolerance for object alignment. The throughput (TP), also known as the index of performance, is calculated as the ratio of the index of difficulty (ID) over movement time:

$$TP = \frac{ID}{MT} \tag{3}$$

What's more, we computed path efficiency (PE) as a measure of kinematic economy:

$$PE = \frac{D}{PT} \tag{4}$$

where PT refers to the path trajectory, calculated as the sum of the differences of the 3D points on the path:

$$PT = \sum_{t=0}^{N} x_{t+1} - x_t \tag{4}$$

## 3.3 Results and Discussion

We collected data from trials of each subject based on the metrics we used. Fitts' kinematic analysis was only computed on successfully completed trials in order to investigate how the AR/VR paradigms influence successful object manipulation motor strategies. In order to account for AR/VR's influence on functional outcomes, traditional metrics (such as task completion rate and total task time) are displayed in Fig. 25.

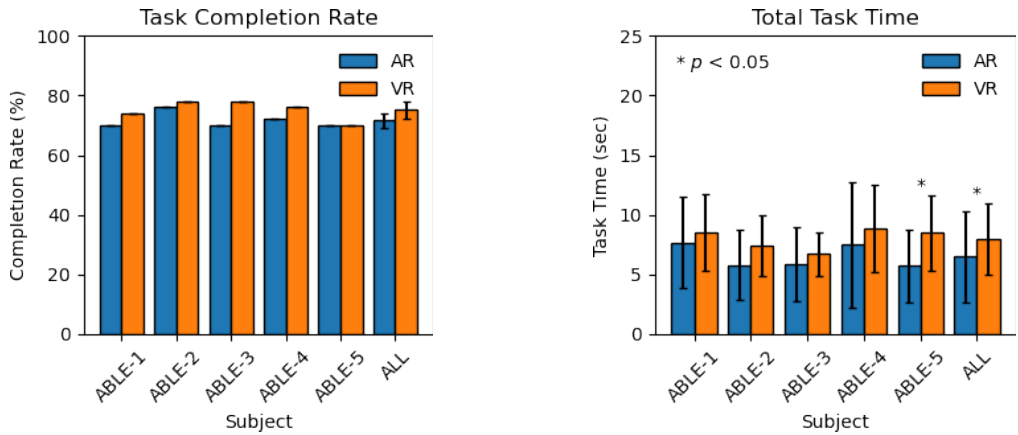

**Figure 25. The task completion rate and averaged task completion time of five able-bodied subjects in AR and VR environment respectively. The bars in the figure represents the mean and standard deviation of the task completion rate and task completion time. A star above the bars denotes a statistically significant increase between the two values (p<0.05). The mean task completion rate shows an increase (+3.6%) in VR. The averaged task completion time shows a statistically significant decrease with p<0.05 in able-bodied subject 5 (-2.782s, -32.81%) and averaged task completion time data from all five subjects (-1.487s, -18.65%) when comparing results in AR to VR (value, percentage).**

In order to investigate whether there was a significant statistical difference of one paradigm over another for the experiments conducted in both environments, we applied a paired-sample t-test on each pair of data generated from both environments by each subject. Incomplete data pairs on each metric were not involved in the paired-sample t-test. The task completion rate on AR and VR tasks for five subjects from 1 to 5 are 70% and 74%, 76% and 78%, 70% and 78%, 72% and 76%, 70% and 70% respectively, which indicates that the task completion rate in both VR and AR environment are quite close to each other, with the task completion rate in VR slightly outperforming that in AR with a small increase of no larger than 8% from each subject and a total increase of +3.6% in averaged task completion rate. The task completion time on AR and VR tasks for five subjects from 1 to 5 are 7.690 ± 3.790s and 8.536 ± 3.201s, 5.802 ± 2.967s and 7.430 ± 2.572s, 5.829 ± 3.107s and 6.710 ± 1.846s, 7.475 ± 3.307s and 8.849 ± 3.653s, 5.699 ± 3.062s and 8.481 ± 3.137s respectively (mean ± standard deviation). The task completion time on AR and VR tasks for the average of all subjects are 6.486 ± 3.828s and 7.974 ± 3.012s. We observe that tasks performed in AR required a smaller amount of time compared with that in VR, with data from able-bodied subject 5 and averaged task completion time reaching a statistically significant decrease ($p<0.05$). Based on the functional characteristics of AR and VR, AR tends to provide participants with stronger cognition of spatial sensation in the immersive scene than that in VR [38]. Furthermore, according to the subjective feedback from subjects involved in this experiment, the visual accessibility of the surroundings from the real-world in AR environment promotes a positive cognitive impact when performing three-dimensional object manipulation tasks in a mixed reality scene, which may partially contribute to the decreased consumption of time compared with the same series of tasks performed in VR environment.

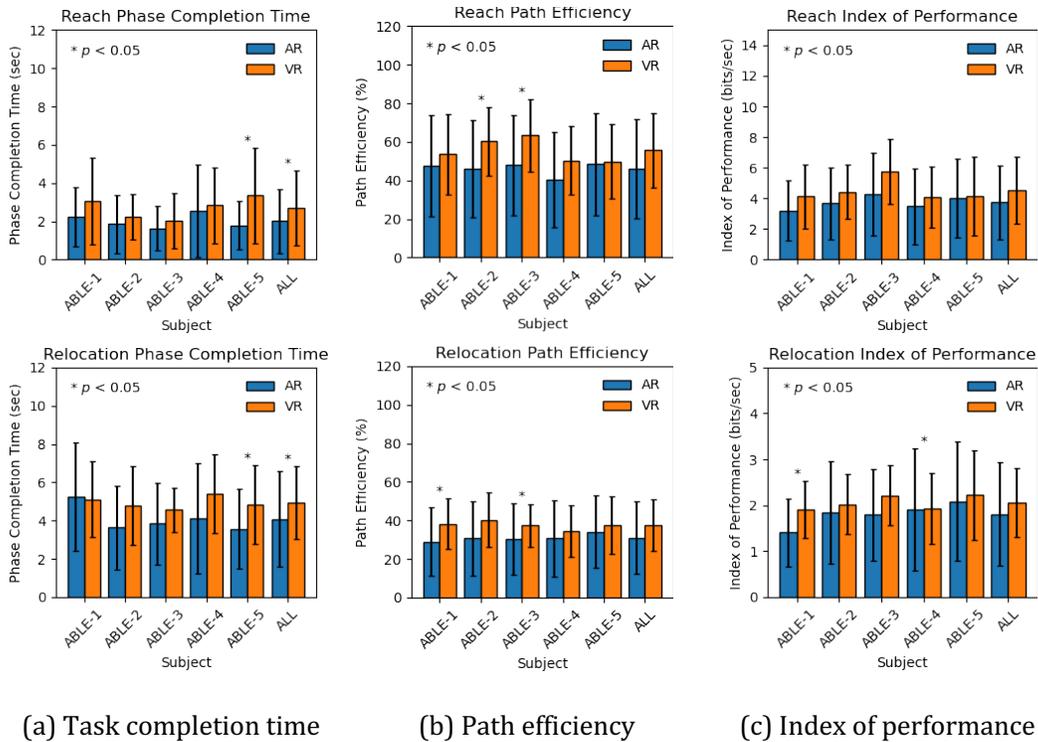

(a) Task completion time     (b) Path efficiency     (c) Index of performance

**Figure 26. The task completion time, path efficiency and index of performance in reach and relocation phases from five able-bodied subjects in AR and VR respectively. A star above the bars denotes a statistically significant increase between the two values (p<0.05). The data of task completion time in AR shows a statistically significant decrease with p<0.05 from able-bodied subject 5 and the average of all subjects in both reach and relocation phases. The averaged path efficiency shows an increase (+9.59%) in the reach phase and (+6.73%) in the relocation phase in VR based on the data from all subjects. The results on the index of performance (Fitts' throughput) shows an increase (+0.783bits/s, +20.94%) in the reach phase and (+0.251bits/s, +21.26%) in the relocation phase in VR based on the data from all subjects (value, percentage).**

In order to lay a comprehensive analysis on the kinematic performance of the virtual prosthetic hand on different motor activities, we computed results for both reach and relocation phase for time consumption, Fitts' throughput and path efficiency respectively shown as Fig. 26(a)-(c) and Table 2-4.

Table 2. Mean and standard deviation value of time consumption (s) from subjects in AR and VR tasks in reach and relocation phases

| Phases of Motion | Reach | | Relocation | |
|---|---|---|---|---|
| Virtual environment | AR | VR | AR | VR |
| Able-bodied subject 1 | 2.223±1.552 | 3.057±2.248 | 5.247±2.810 | 5.097±1.985 |
| Able-bodied subject 2 | 1.860±1.524 | 3.001±2.337 | 3.629±2.190 | 4.798±2.050 |
| Able-bodied subject 3 | 1.642±1.160 | 2.026±1.434 | 3.832±2.123 | 4.555±1.154 |
| Able-bodied subject 4 | 2.544±1.401 | 2.834±1.589 | 4.107±2.899 | 5.377±2.059 |
| Able-bodied subject 5 | 1.793±1.251 | 3.353±2.500 | 3.559±2.096 | 4.856±2.063 |
| Total | 2.013±1.659 | 2.681±1.962 | 4.069±2.500 | 4.934±1.890 |

Table 3. Mean and standard deviation value of averaged path efficiency (%) from subjects in AR and VR tasks in reach and relocation phases

| Phases of Motion | Reach | | Relocation | |
|---|---|---|---|---|
| Virtual environment | AR | VR | AR | VR |
| Able-bodied subject 1 | 47.48±26.14 | 53.66±20.77 | 28.96±17.88 | 38.29±13.28 |
| Able-bodied subject 2 | 46.05±25.00 | 60.27±17.86 | 30.63±19.26 | 40.14±14.15 |
| Able-bodied subject 3 | 47.88±26.12 | 63.34±18.83 | 30.12±18.57 | 37.44±11.05 |
| Able-bodied subject 4 | 40.27±24.68 | 50.28±17.81 | 30.62±19.81 | 34.62±13.47 |
| Able-bodied subject 5 | 48.36±26.55 | 49.72±19.24 | 34.06±18.88 | 37.65±14.88 |
| Total | 45.97±25.63 | 55.56±19.49 | 30.89±18.79 | 37.62±13.38 |

Table 4. Mean and standard deviation value of Fitts' throughput (bit/s) from subjects in AR and VR tasks in reach and relocation phases

| Phases of Motion | Reach | | Relocation | |
|---|---|---|---|---|
| Virtual environment | AR | VR | AR | VR |
| Able-bodied subject 1 | 3.205±1.955 | 4.136±2.098 | 1.409±0.742 | 1.910±0.627 |
| Able-bodied subject 2 | 3.676±2.350 | 4.426±1.770 | 1.849±1.115 | 2.025±0.663 |
| Able-bodied subject 3 | 4.292±2.700 | 5.766±2.137 | 1.792±1.004 | 2.216±0.655 |
| Able-bodied subject 4 | 3.493±2.476 | 4.071±1.995 | 1.909±1.328 | 1.928±0.776 |
| Able-bodied subject 5 | 4.019±2.556 | 4.158±2.564 | 2.088±1.295 | 2.225±0.973 |
| Total | 3.739±2.431 | 4.522±2.197 | 1.811±1.130 | 2.062±0.751 |

According to the results, the mean time consumed in AR is less than that in VR from all subjects in both reach and relocation phases with a statistically significant decrease (p<0.05). This is in accordance with the trend of data found in total task completion time. Furthermore, we discovered a general increase from each subject for mean path efficiency in VR when compared to AR, with the mean path efficiency value in VR from subjects 2 and 3 reaching a statistically significant increase (p<0.05) in the reach phase and subjects 1 and 3 reaching a statistically significant increase (p<0.05) in the relocation phase. The averaged path efficiency value from all subjects in the reach phase in VR has an increase of +9.59% when compared to AR. Such results demonstrate that the trajectory path of the prosthetic hand is less variant and closer to the straight-line path between the initial hand location and the location of the target holder, which illustrates that subjects tend to achieve better control regarding the adjustment of the spatial position of the virtual prosthetic hand in VR than that in AR. Furthermore, since the training scene simulated by VR provides a virtual room and avatar that represents the body of a subject in the virtual scene, geometric primitive objects appear to be more consistent on their peripheral scale with the target holders of the virtual PHAM frame in VR after subjects get familiar with the virtual room in VR. Therefore, subjects were able to adapt a shorter and less variant path trajectory in the VR environment.

Moreover, the Fitts' throughput (index of performance) in VR on average outperforms that in AR in both reach and relocation phases (+0.783bits/s, +20.94% and +0.251bits/s, +21.26%), with a statistically significant increase for subjects 1 and 4 during the relocation phase ($p<0.05$), which illustrates that subjects tend to spend less time on more complicated object manipulation tasks in VR than in AR.

Comparing the data of the three metrics across phases of motion, we discovered that subjects achieved better performance regarding all aspects--time, path efficiency and throughput--during the reach phase than that during the relocation phase. This is partially due to the lack of needs to control the virtual prosthetic hand to perform grips during the reach phase, which saves more labor and mental effort when controlling the movement of the virtual arm. Another potential factor that might contribute to this phenomenon is the decrease of attention and muscle strength as a task was carrying along, which may lead to a subtle impact based on the chronological order of the two phases. This can be further investigated in the future by subdividing the relocation phase into a relocation and return phase, wherein after the object is transferred to the destination holder, the return phase is triggered. The return phase does not require any control of the virtual prosthesis but chronologically succeeds the relocation phase. With that being investigated, we can determine whether subjects' performance on controlling the virtual prosthesis deteriorates with the chronological order of phases of motion. Moreover, the increased accessibility of VR technologies when compared to AR technologies for subjects in their daily lives might also cause differences in these results. However, we did not analyze any behavioral habits further in this work since this experiment focused on comparing the efficacy and degree of assistance provided by VR and AR in prosthetic manipulation.

## 3.4 Conclusion

This paper presents a study focusing on the comparison between augmented reality and virtual reality on their efficacy in assisting prosthetic control. We implemented a human- computer interactive prosthetic control system that is capable of facilitating users with real-time myoelectrical prosthetic control practice through performing dexterous motor tasks in virtual and augmented reality based on PHAM. A pilot experiment was conducted with five able-bodied subjects performing object grasping and relocation tasks in both VR and AR under the same protocol. An assessment approach was applied based on task completion and Fitts' kinematics analysis to quantitatively evaluate the kinematic performance of subjects in both paradigms during the experiment. The results showed an advantage of VR over AR in assisting subjects to achieve motor tasks with higher path efficiency, throughput and larger likelihood of successful completion, whereas AR outperforms VR with less time consumed for task completion. In the future, we would like to further investigate whether the motor skills learned in augmented or virtual reality transfer to increased motor proficiency in physical reality.

# Chapter 4. MyoTrain: An Augmented Reality-assisted integrative system for Upper-limb prosthesis control

In this chapter, we introduce an integrative AR training system—MyoTrain system for upper-limb prosthesis control. The aim of the study is to further explore the effectiveness and the feasibility of AR-assisted prosthesis control system in comparison with those by a physical prosthesis control system following the same function mechanism and standard, and what further enhancement or inclusion of other paradigms can facilitate the AR system with better performance.

## 4.1 Background introduction

Upper limb motor deficits, typically caused by neurological disorders or loss of limbs, encumbers people with constraints and limitations on tackling motor tasks happened in everyday life, restraining their independency on achieving activities of daily living. In consequence, the constraints and inconveniences brought up by motor deficits may generate a negative impact on a victim's daily life both physiologically and psychologically and therefore a serious issue that needs attention. Current treatment involves prostheses, an artificial device that assist people with motor deficits restore functions and mobility, which help them achieve better performances when dealing with motor tasks and activities at work or in daily life. Early stages of prostheses for amputees or people with needs include body-powered prostheses, a traditional type of prosthetic device that is typically operated by user's own body through a harness and a cable through mechanical principles. And in recent decades, the evolution of prosthetic devices brings prosthetic users with more senses of information from different dimensions, which improves the functional manipulations of prostheses and enhances user experiences. An electrical prosthesis powered by myoelectric control can help users perform more motor activities accurately and diversely. When operating the device, electrodes that are attached to users' residual limbs acquire input signals from muscles, which are then processed and decoded into specific commands to manipulate the prosthesis [38][39]. In order to operate a myoelectric prosthesis to perform dexterous tasks smoothly, users need to practice the contractions and flexions of target muscle groups for better control, which is known as prosthesis-fitting. In order to better fit a myoelectric prosthesis, subjects typically will undergo a training program in which an occupational therapist is involved to instruct subjects with practice of muscle activation and deactivation for prosthesis control. Such process needs participants to focus their attention for a considerable amount of time, which are challenging for people to maintain. And the lack of engagement and motivation is another factor that refrains participants from accomplishing the training [24].

With the advent and maturity of augmented reality and virtual reality, researchers and bioinstrumentation developers have tried to incorporate these paradigms into prosthesis fitting and control in recent decades. As mentioned earlier in previous chapters regarding the advantages of VR and AR assistive training systems, more level of engagement and motivation brought by the system can effectively alleviate the monotony and depression during the process. And moreover, the portability of such systems and the flexibility of adjustment on the display content make it more convenient for users to practice at home based on their individual needs.

Compared to virtual reality, an augmented reality-rendered immersive scene is more realistic and resonates more with human's visual preferences. Different from the scenes rendered by virtual reality in which users are exposed to a fully simulated virtual world, augmented reality technology overlays virtual objects onto the background in the real-world. Users can see the simulated object as well as their real-world surroundings at the same time, which augments the sense of presence and reduces the ambiguity of the place where they are located. Some studies have delved into the development and assessment of a prosthesis training or control system with the incorporation of augmented reality [22][93][115][116]. But currently according to author's review and best of his knowledge, many literatures and studies lay more emphasis on validating the feasibility and demonstrating the effectiveness of the AR-facilitated system, and inconsistency and ambiguity lay on the performance and the quantified effectiveness of AR system in assisting prosthesis training and control compared to a physical prosthesis system.

Therefore, this study aims to look into the effectiveness and the performance of a AR-based system compared to a physical reality-based system in assisting prosthesis control and see what specific tools and add-ons can be equipped to improve the effect of the AR system. The primary goal of this work is to identify strategies to enhance the AR system to better match the performance in the real-world. Throughout the experiment, we aim to identify to what extent AR-based prosthetic control system may potentially lags behind manipulation of prosthetic hand in reality on task performance and come up with solutions to address the existing limitations and improve the effectiveness and efficacy of operating AR-based prosthetic hand to perform tasks as a consequence. Moreover, we also aim to identify if weighting the limb during AR training is necessary to better match real-world task behavior. If there is no major difference in task performance with the bypass prosthesis and the task performance in AR can match the real-world task performance, then it is possible that training in augmented reality with a wireless armband alone could result in performance improvements in the real-world. However, if there is an effect due to the weight of the bypass prosthesis on the participants arm, then pre-prosthetic interfaces should be similarly weighted to real-world prosthesis to achieve better results.

## 4.2 Methods

### 4.2.1 System Design

Aiming to quantitatively compare the effectiveness of AR-based system with that of physical reality-based system in facilitating prothesis control and see if an addition of an bypass prosthesis to the AR system can make an improvement to the AR system, we built three set of prosthesis control environments: physical reality-based system with a bypass prosthesis (Phy-BP), augmented reality-based system without a bypass prosthesis(AR-null), and an augmented reality-based system facilitated with a bypass prosthesis (AR-BP) to investigate the potential difference on boosting the effect of prosthesis control between a physical system and an AR system and see if a bypass prosthesis could narrow the gap. We control the variability of all other potential features and factors that might generate a bias towards any of the three systems, which means that, apart from the only difference on system-level component in physical or AR environment and the equipment condition of the bypass prosthesis based on the needs for the experiment, all other features, including the protocol underwent in each environment, the mechanism to control the prosthesis, and the tasks that subjects need to accomplish in the three systems, are all kept the same.

The component of the system can be illustrated based on the environment they are in accordingly: all three systems applied the assessment tasks from Prosthetic Hand Assessment Measure (PHAM) [95]. In physical environment, the system is composed of a PHAM frame, with the target object cylinder, the bypass prosthesis arm and the corresponding adjunct sytems, such as the Vive® tracker (HTC Corp., Xindian, Taiwan) applied for motion tracking, and the Myoband® (Thalmic Labs, Ontario, Canada) applied for myoelectric signal sampling. And the pattern recognition algorithm: RESCU developed by Infinite Biomedical Technologies (Baltimore, USA) is applied for sEMG signal decoding in the three environments. For AR-null and AR-BP environment, besides all these control system and motion tracking hardwares, Vive® Pro Eyes Headset (HTC Corp., Xindian, Taiwan) is used for providng the AR visual feedback. What's more, for the bypass prosthesis, besides the bypass framework in joint collaboration with BLINC Lab in University of Alberta, we applied the BeBionic Hand 3 (Ottobock, Germany) to perform motor activities such as grasp and wrist rotation. Meanhwhile, for the software component, Python 3.7.4 (Python Software Foundation, Dover, DE), and Unity3D® 2018.4.17 (Unity Technologies, San Francisco, CA) was installed and setup on the desktop for carrying out this experiment. We used a computer with i7-10750H CPU@2.60GHz, 16GB RAM, and NVIDIA GeForce® RTX 2060 graphics along with Windows® 10 operating system to run the software system. Fig. 27 gives an intuitive overview of the three systems and their components in detail. Fig. 28 gives a detailed description of the systematic component of the MyoTrain AR system and the way it functions.

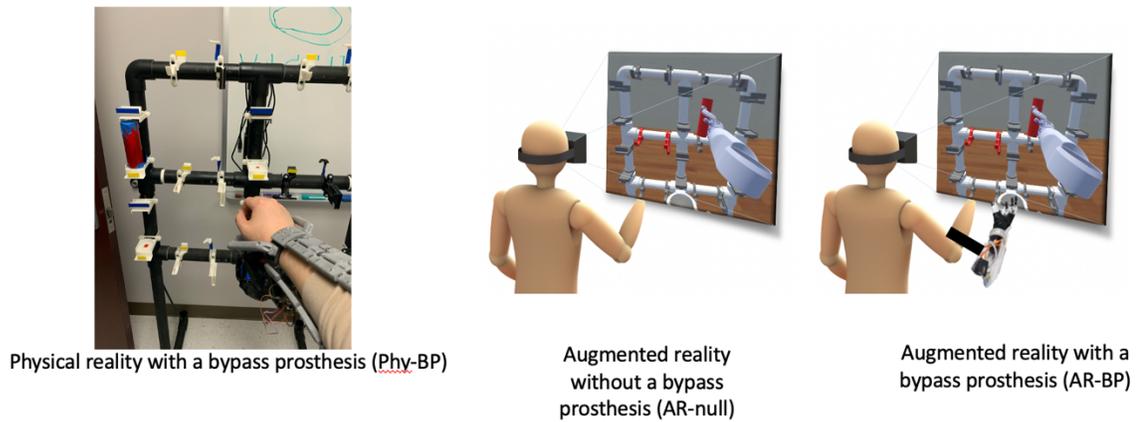

Physical reality with a bypass prosthesis (Phy-BP)　　Augmented reality without a bypass prosthesis (AR-null)　　Augmented reality with a bypass prosthesis (AR-BP)

**Figure 27.** An overview of the three experiment environments. In order to quantitatively assess and compare the physical reality-based system and AR-based systems along with the potential impact a bypass prosthesis will make on the AR system, the three experimental environments: Phy-BP, AR-null and AR-BP are built.

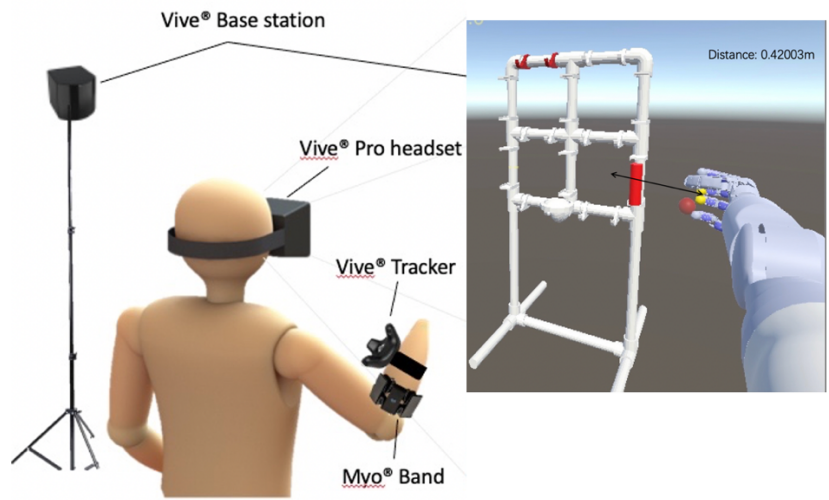

**Figure 28.** The MyoTrain AR prosthetic control system. The functional outcome measurement—PHAM is virtualized in both an AR environment. A Vive® tracker is attached to the right arm of a subject to track the position and the motion of the subject's arm in real-time; A Myo®band is placed next to the Vive® tracker to collect the sEMG data

**from the muscles and transmitted to the computer, which is then decoded into control commands to activate different grasps of the virtual prosthesis. The virtual prosthesis along with the PHAM frame will then be dynamically simulated in AR and displayed in a Vive® Pro headset as visual feedback to the subject as a consequence.**

### 4.2.2 Experiment Protocol

In this study, we recruited two able-bodied subjects (25±2 yr) into this experiment. Both subjects were right-handed and all manipulation tasks in the experiment were achieved by subjects through their dominant hand. The experiment was conducted with their informed consent. Before going through the experiment, we acknowledged subjects the overall experiment protocol. This experiment protocol was approved by the Johns Hopkins Medicine Institutional Review Board (JHM IRB). This protocol includes the experimental procedures that each subject goes through, along with the metrics used for evaluating the kinematic performances of prosthesis control.

The hypothesis of this experiment is that the initial reaching and relocation phases should be pretty similar between the real-world and AR tasks. However, in the reaching phase, we assume a longer movement time consumed in AR environment. As the subject approaches the target location, the subject may slow the arm considerably earlier with the virtual arm than the real arm. This could be manifested in a lower yet earlier peak velocity for the virtual task due to the uncertainty of the arm's position relative to the target in the virtual environment. If the subjects struggle with placing their hand in the correct location, we should also anticipate seeing a less efficient path. We assume that the time consumed in the relocation phase in AR and physical environment may be quite close to each other, with that in AR a bit more than that in physical environment. We also anticipate that the trajectory of the virtual hand in AR environment in all phases may spans a longer distance and variant path, which will result in a lower path efficiency in both reach and relocation phases.

**4.2.2.1 Object Manipulation Tasks**

In this experiment, we will compare three conditions: 1) Real-world task completion with a bypass prosthesis on the Prosthetic Hand Assessment Method (PHAM), 2) AR task completion with a bypass prosthesis on the virtual PHAM, and 3) AR task completion without the bypass prosthesis on the virtual PHAM. When doing tasks in the physical environment, the bypass prosthesis will be controlled using EMG-based pattern recognition from Python to interact with the physical objects. In the AR task with the bypass prosthesis, a virtual arm will be overlaid on the bypass prosthesis and the virtual arm will be operated with EMG-based pattern recognition from Python. In this condition, the bypass prosthesis will be inactive. In the AR task without a bypass prosthesis worn by the subject, a virtual arm will be positioned as attached to their right intact arm and the virtual arm will be operated with EMG-based pattern recognition from Python. During the setup for the AR tasks, the participant will be moved to an area clear from obstacles and the virtual PHAM will be oriented on the floor in front of the subject.

**4.2.2.2 Segmentation of Phases**

In order to better compare the performances conducted in three environments and understand where there are differences between the physical reality and AR, we will segment the PHAM task into different phases. A single trial of the AR PHAM and real PHAM will be divided into 3 phases: 1) Reach Phase, 2) Relocation Phase and 3) Return Phase. In the virtual task, this can be triggered when the start/stop button is pressed, when the object is contacted/lifted, and when the object is dropped using virtual markers. In the real task, we're using a digital switch to initiate the PHAM task trial, which can be monitored. To mark when the phase transitions from the Reach to Grasp or from Relocation to Release phases, we can measure when contact is first initiated with the target object using force sensitive resistors (FSRs) attached to the object. Once the object is grasped and lifted by the prosthetic hand, the force will be dropped down to 0 and when placed on the target holder, the force will arise to the anticipated level.

**4.2.2.3 Evaluation Metrics**

After the tasks are performed by subjects and data are obtained, we will assess the system in assisting prosthesis control based on these metrics that reflect subjects' kinematic performance and smoothness in performing these tasks: A) Path efficiency, B) time consumed in each segment, and C) xz-coordinate trajectory plot.

**4.2.2.4 Experiment procedures**

In this experiment, each one of the subjects will go through the manipulation of the system to perform 4 different object manipulation tasks separately in Phy-BP, AR-null and AR-BP. The order they go through the three environments is randomized for unbiased trials in each testing environment. The four different object manipulation tasks are specially designed as shown in Fig. 29. Subjects will perform each type of tasks for 30 times consecutively and will get 10 min rest between two different types of tasks to relax their muscles. After completing all trials with the first task type, the subject should switch over to the alternative task type. Subjects are allowed to practice the tasks for 30 min prior to the start of data recording.

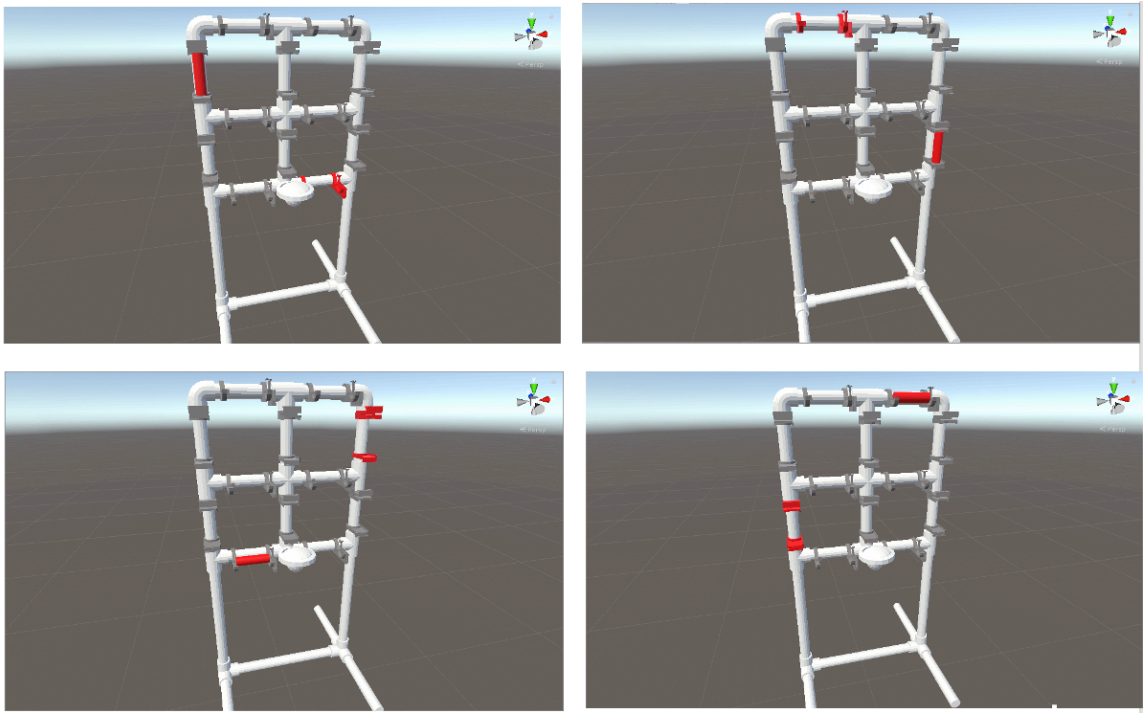

**Figure 29. Four different types of motor tasks. The pictures above indicate the four specific types of tasks a subject will go through. The distance between the initial and destination locations are kept the same among the four tasks. Subjects will repeat the trials of each task 30 times and go through the four different types of tasks in a random order.**

A) *Myo®band Preparation*

Before going through the experiment, subjects will need to acknowledge the overall protocol, then when they are ready, they will don the Myo®band at the position of the right forearm with maximum muscle mass.

B) *Pattern Recognition Algorithm Training*

After the Myo®band is worn properly, researchers will launch python scripts for training the pattern recognition (PR) algorithm and instruct subjects to go through the training phase. First, subjects will need to follow researchers' instructions to make sure the Myo®band is connected to the PC. Next, subjects will need to go through the standard PR training phase. On a computer screen, various cues on performing each of the motor activities (hand open, hand rest, power grasp, wrist pronation and wrist supination) will be prompted (via cue images) in succession. Each cue will be prompted for 5 seconds, with the last 3 seconds worth of EMG data being collected. The succession of cues will be completed 3 times in 3 different arm positions: (1) at the side with a 90° elbow bend, (2) outstretched in front of the body at a comfortable posture, and (3) outstretched in front of the body and raised above the head.

If accuracy results for any of the trained patterns are less than 80%, then the training phase should be restarted to ensure the stability of the myoelectric control. Everytime when users have taken off the Myo®band and don it again because of muscle rest, they need to go through the PR training process again.

C) *Device Preparation*

Before going into the sampling phase, the space for conducting the experiment needs to be set up. For the AR PHAM setup, it is assured that the scanned area by two base stations will cover the area where all the devices (Vive® Pro headset, two Vive® trackers) might potentially reach during the experiment. Then a Vive® tracker for locating the PHAM frame will be placed on the ground, and subjects will wear the Vive® Pro headset on their head and don a Vive® tracker on their right forearm next to the Myo®band. Before starting the sampling phase, subjects and researchers run the system quickly and adjust the positions of the two Vive® trackers to ensure the PHAM frame is placed at a proper position with its front facing straight to subjects, and the virtual arm overlaying on the bypass prosthetic hand or the intact arm in the AR scene. For the physical PHAM setup, force sensitive resistor tapes will be attached to the surface of the four geometric primitives and the contact surface of holders. The BeBionic® hand will be attached to the bypass prosthesis and researchers check the operational conditions of the prosthetic hand to make sure that it's able to perform motor activities properly. Additionally, the FSRs attached to the geometric primitives are checked to ensure it functions correctly.

D) *Pre-training Phase*

In this phase, subjects will be required to go through the manipulation tasks in Phy-BP, AR-null and AR-BP. Considering the potential impact that any individual prior experiences in AR might bring, we attempt to pre-train every subject in all three training environments in order to familiarize them the environments and minimize the potential empirical impact. Subjects will go through 30 trials in these three environments prior to the experiment. After finishing each session, subjects will be given 10 minutes to relax their muscle groups.

E) *Experiment Phase*

After going through the pre-training phase and allowing enough rest following the end of the phase, when subjects are ready, they can carry along into the experiment phase. Subjects will go through the 4 sequential tasks, with each task repeating 30 times to relocate the geometric primitives as cued by the system in Phy-BP, AR-null and AR-BP respectively. After each sequential task is finished in any environment, subjects will be given 10 min for a break, and when continuing back to the other group, subjects will go through the pattern recognition training phase again to retrain the Myo®band. After repetitions of a single task in a group is finished, subjects may take a break, approximately 10 min to relax their muscle groups. For a single trial of object manipulation, the subject will complete the following:

1. The initial location of geometric primitive and the target location on the PHAM frame will be highlighted to users. Press the button of the PHAM frame to signal the start of the manipulation task. The timer will start counting the time and each trial will give subjects 30 secs to do the manipulation task.

2. Control the virtual prosthesis or bypass prosthetic hand to perform a series of reach, grasp, and relocation, release and return tasks to transport the prompted geometric primitive to the target holder.

3. Touch the button of the PHAM frame once again to signal the end of the manipulation task. This signals the end of a successful task.

4. If the object manipulation cannot be completed within 30 sec or if the object is dropped to the ground, the task will immediately end and recorded as failed, and the next manipulation task will be loaded.

5. Keep repeating the same task for 30 times

In total, each subject performs 30x4 trials in the three environments.

## 4.3 Results and Discussion

After having (N=2) subjects going through all trials, we obtained the data based on our evaluation metrics and conducted a statistical analysis to reflect the average performance of each subject in each type of task respectively. Fig. 30(a) and (b) shows the plot of the averaged path efficiency along with the variation of the data by subjects in each phase of the experiment. Fig. 31(a)-(f) shows the average of time consumed in each phase by subjects.

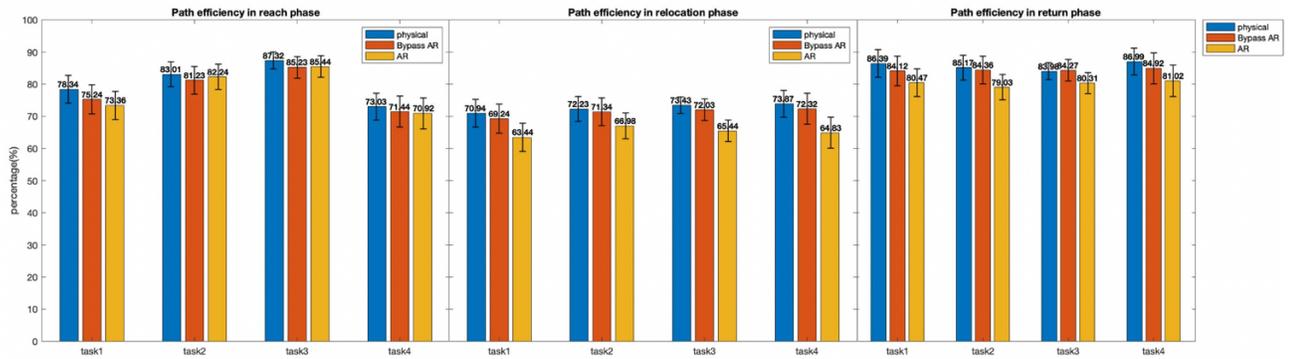

(a) Subject 1

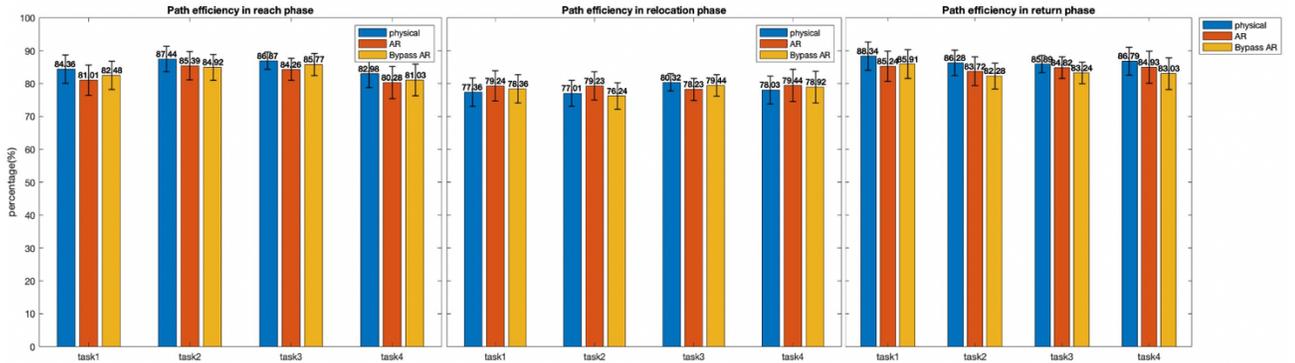

(b) Subject 2

**Figure 30. The averaged path efficiency in each task by subjects. The data on each bar represents the averaged path efficiency value of 30 trials on each task by the two subjects. The error bars denote the standard error of mean.**

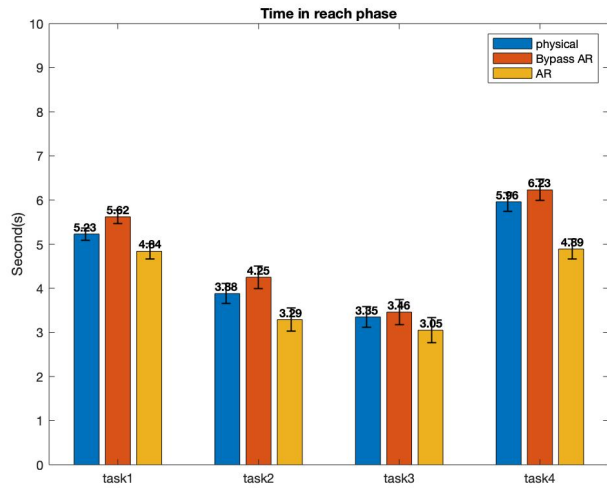

(a) Subject 1

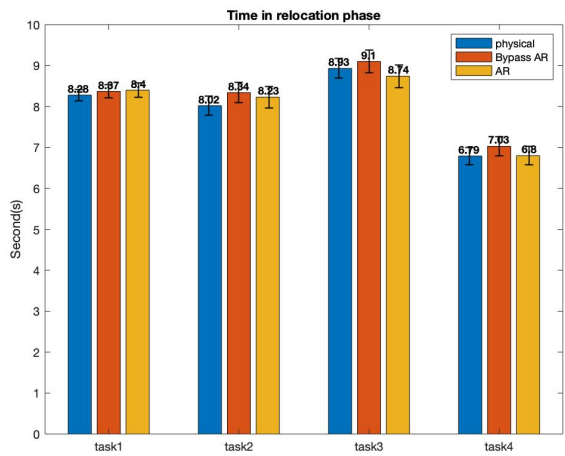

(b) Subject 1

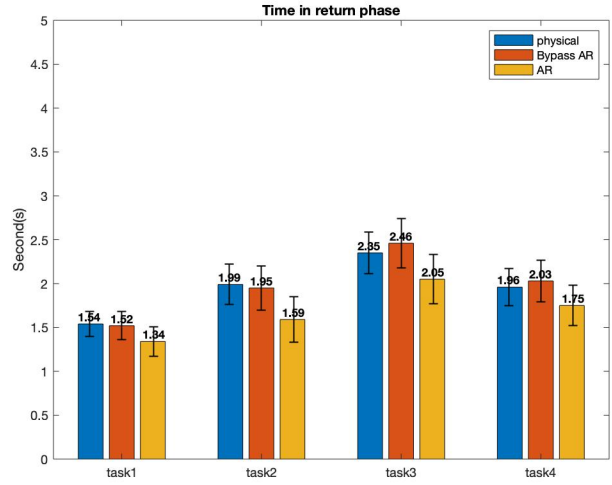

(c) Subject 1

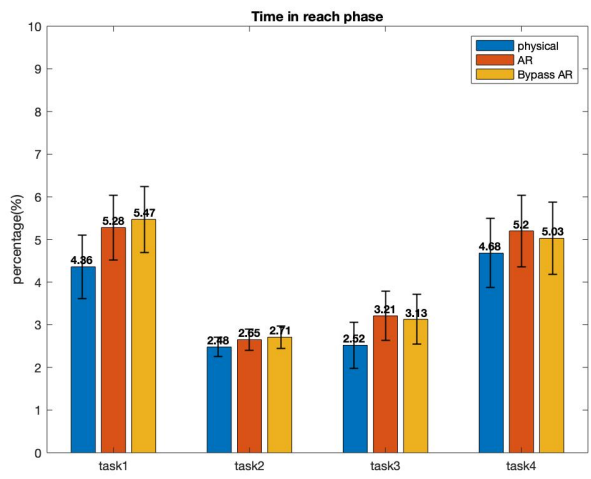

(d) Subject 2

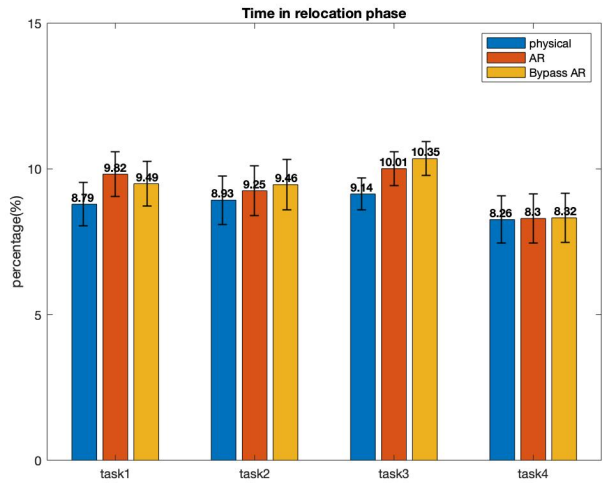

(e) Subject 2

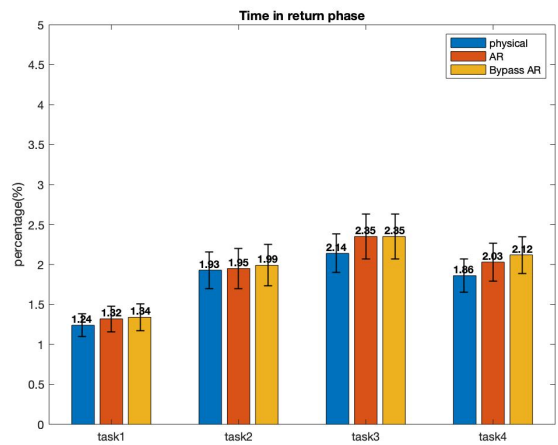

(f) Subject 2

**Figure 31. Average time consumed in each phase. (a)-(c) are the plots of average time consumed in reach, relocation and return phase by able-bodied subject 1. (d)-(f) are the plots of average time consumed in reach, relocation and return phase by able-bodied subject 2.**

From Fig. 30 and 31 shown above, we can first observe the trend from horizontal comparison that among all three phases, the path efficiency in the relocation phase is the highest from both subjects and meanwhile, the average time consumed in the relocation phase is the shortest. This is because that compared to reach and return phase, the distance that the arm needs to travel along is significantly longer in the relocation phase. What's more, subjects need to maintain their control of the prosthesis arm in 'grasp' pattern and keep the object from potentially dropping to the ground due to an interruption to balance. In addition, in each of the four tasks, the average path efficiency in Phy-BP by two subjects achieve modestly higher than the other two environments in general. Yet in the relocation phase in which more motion activities are conducted, the average path efficiency of AR, especially by subject 1, perform a lower value than the other two. And the path efficiency by AR-BP, the AR system equipped with a bypass prosthesis, is higher than AR-null and reached close to that in the physical environment from both subjects. This demonstrated that by equipping a bypass prosthesis in AR system, users can achieve similar path efficiency compared to physical reality system. However, when looking at the data in the average time consumed in different tasks, we observe that there's a slight decrease in Phy-BP system in all phases from a general perspective, and the AR-BP environment didn't show trend in alleviating the consumed time on average in different phases. This may be caused by the weight allocated to the intact limb of subjects that apply more burdens to the movement compared to AR-null system, which to some extent, counteracts the encouraging effects provided by the visual cues from the bypass prosthesis. Fig. 32 shows the trajectory plot on xz-coordinate (front view plane).

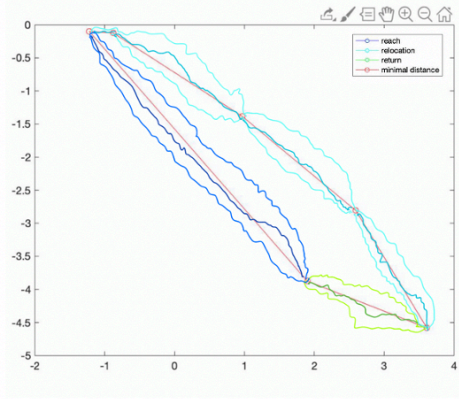
Phy-BP

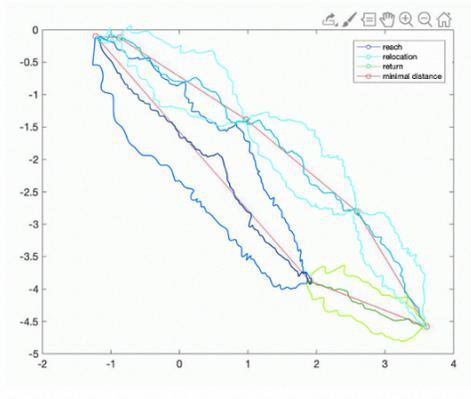
AR-null

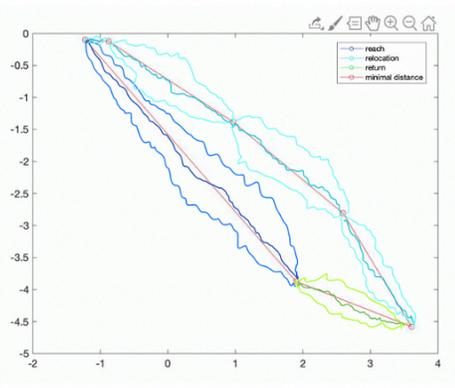
AR-BP

(a) Subject 1

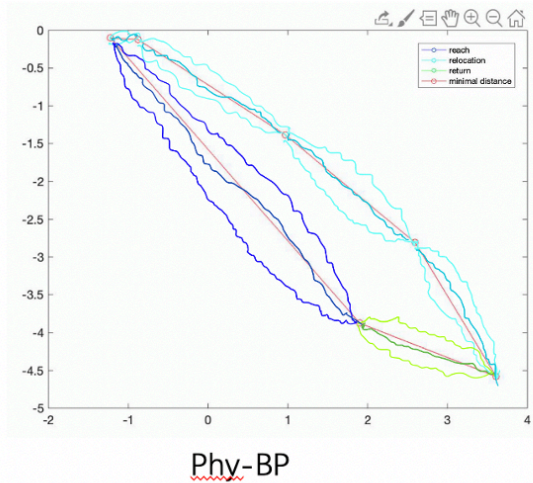
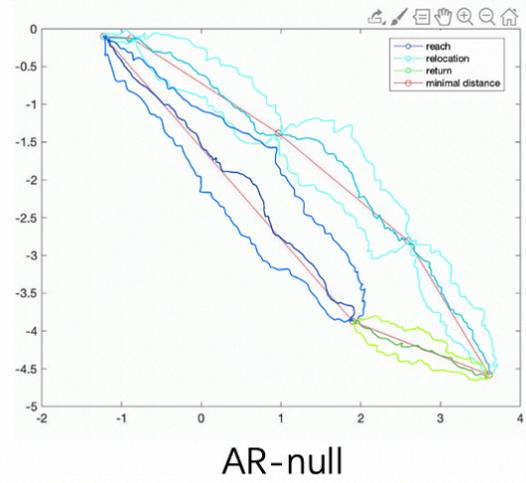
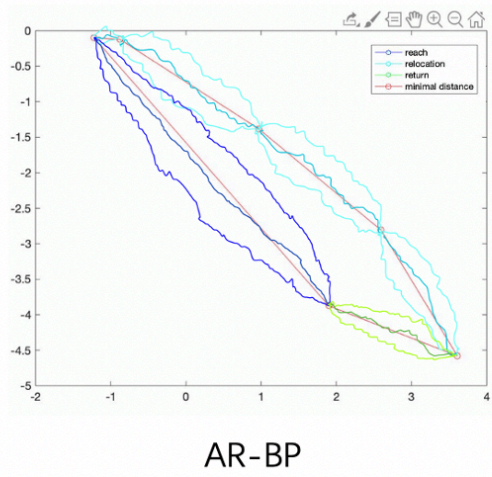

(b) Subject 2

**Figure 32. xz-coordinate trajectory plot. Fig. 32(a) and (b) shows the trajectory plots of subject 1 and 2 in performing task 1. The blue, cyan and green colored lines represent the trajectories in different phases. The darker lines are the averaged trajectories whereas the lighter lines are the outer range of the trajectories.**

From Fig. 32(a) and (b), we observe that when performing the same tasks, the outer range of the trajectories in AR-null are much larger than that in Phy-BP and AR-BP environment based on the plots from both subjects, which means that when performing tasks in AR environment without a bypass prosthesis, the movement of the prosthetic arm is more off the optimal trajectory compared to the other two environments. And we see that with the equipment of a bypass prosthesis, the outer range converges and getting closer to the range of Phy-BP in comparison, which means that the equipment of a bypass prosthesis in AR system can keep the motion of the prosthesis arm more stable and less fluctuating.

## 4.4 Conclusion

In this study, we built up an integrated augmented reality-facilitated system—MyoTrain to quantitatively evaluate and compare the effectiveness of AR prosthesis control system compared to physical reality-based system. We conducted a pilot experiment to study the potential difference of effects between AR and physical reality-based system and involve the use of a bypass prosthetic arm to facilitate the AR system. Based on the average improvement in average path efficiency and converged outer range from xz-coordinate trajectory, we demonstrated the feasibility of the MyoTrain system and found the subtle deficiency in path efficiency, cost time and fluctuation of movement in AR system compared to the physical reality system, and conclude that by equipping a bypass prosthesis, the performance in path efficiency and fluctuation of trajectories are enhanced. Future work will involve more subjects to gather more data to firm up this conclusion and look into other techniques such as eye tracking that can potentially improve the effectiveness of an AR system.

# Chapter 5. Future Directions

In this chapter, we will give an overview of the current challenges and limitations of VR and AR assistive applications in the rehabilitation field, and give assumptions and suggestions on how these issues can be potentially addressed and further improvements can be made.

The application of VR/AR technology into rehabilitation treatment brings benefits to patients and therapists. However, there are still challenges and limitations that are in need to overcome to make this technology better serves rehabilitation area. One of the aspects that rises concerns is the design and personalization for the content of a rehabilitation program that would fit individuals properly. For example, regarding stroke rehabilitation, exercises should be given occupational focus and contextual relevance to the user's personal life and needs. Many of the games or training environments currently designed are generally rote exercises, lacking in personalization for the user; tasks that are specific to the home should be developed to allow users to regain their independence at home [103]. However, allowing for that level of customization for a patient population with highly variable needs brings with it challenges of development scalability, tractability, and support that are not easily dealt with and would consume financial and technical effort and investment. What's more, some virtual reality assistive tools typically lay requirements on certain amount of space where users conduct training or treatment, which raise concerns for some groups of users who want to conduct such trials at home. In addition, human-computer interface performance is still highly variable, indicating the need for affordable affiliated products and equipment, such as motion-tracking systems with improved accuracy of movement detection, or high-resolution screen or lens through which the imaging quality of the projected scenes in VR or AR can be greatly enhanced and closer to the comfort range of human's retina.

Furthermore, some patient populations, especially those who have limited access to virtual game or interaction experiences, may find themselves hard to be accustomed to the immersive experiences provided by VR/AR, which may result in a vulnerability to health-related side effects such as dizziness, headache, and fatigue [104]. Such problems should bring more attention when looking into the VR/AR applications in the rehabilitation area where senior populations tend to take a large percentage in some branches and are more likely to be less experienced with the use of virtual reality-related services. Compared to immersive VR display such as those systems with head mounted display from which users can hardly get visual exposure and acknowledgement to real-world, AR display or non-immersive VR display which typically uses flat monitors or screens tend to work better and less likely to cause nausea, dizziness or fatigue since the view of real-world surroundings are available in these paradigms, which presents users with more spatial and locational references [105]. However, due to the naturalness and the necessity of some designs that would require a VR-based environment, the degree of immersion of the display to users for better visual effect without causing discomfort to some populations would be a tradeoff.

Providing higher sense of presence to users by engaging multi-degree of sense of feelings, such as audio, somatosensory, and olfactory feedback, can enhance the level of reality to users for better simulation and immersive experiences and meanwhile, alleviate the sense of ambiguity and discomfort of users when their eyesight and attention are more attracted by virtual environment and not getting accustomed to it.

Specifically, ways to improve users' experiences in immersive virtual environments include stereovisions accompanied with stereophony audio feedback to present users with a sense of feeling towards their spatial locations relative to the virtual environment they are engaged in. Another direction on facilitating sense of presence in virtual environments can be focused on boosting the quality of interactivities between users and virtual environments. For instance, by simulating the sense of touch in the real-world and providing such levels of feedback to users, an inclusion of real-time sensory or haptic feedback while interacting with a virtual object can enhance the immersive experiences and narrow the gap between reality and the virtual world. The implementation of incorporating such levels of feedback can be improved by combining and applying the state-of-the-art techniques in this area, including vibrotactile wearable devices such as cyber gloves, or encountered-type haptic devices such as quadcopter to provide haptic and force feedback [106]. And depending on the development of new tools and methods in such area, more efficient and user-oriented VR/AR system can be made available and fit clinical needs properly by integrating these cutting-edge techniques.

Finally, more clinical trials need to be conducted to validate the efficacy of VR/AR assisted rehabilitation systems as replacements for traditional therapies. Even though many evidences regarding the efficiency and the effectiveness of VR/AR assistive tools for rehabilitation purpose can be sufficiently found in existing literatures, many of these are conducted based on experimental trials at research or scientific institutes. Larger sample sizes upon trials and tests in clinical treatment and diagnosis are in need to sufficiently demonstrate and promote the power of such paradigms in rehabilitation and expose potential flaws for researchers and tool developers in this area to tackle and improve.

[38] Murray, C.D., Patchick, E., Pettifer, S., Caillette, F. and Howard, T., 2006. Immersive virtual reality as a rehabilitative technology for phantom limb experience: a protocol. CyberPsychology & Behavior, 9(2), pp.167-170.

[39] Lamounier, E., Lopes, K., Cardoso, A., Andrade, A. and Soares, A., 2010. On the use of virtual and augmented reality for upper limb prostheses training and simulation. 2010 Annual International Conference of the IEEE Engineering in Medicine and Biology. IEEE, pp. 2451-2454.

[40] Lambrecht, J.M., Pulliam, C.L. and Kirsch, R.F., 2011. Virtual reality environment for simulating tasks with a myoelectric prosthesis: an assessment and training tool. Journal of prosthetics and orthotics: JPO, 23(2), p.89.

[41] Belter, J.T., Segil, J.L. and SM, B., 2013. Mechanical design and performance specifications of anthropomorphic prosthetic hands: a review. Journal of rehabilitation research and development, 50(5), p.599.

[42] Johansson, B.B., 2000. Brain plasticity and stroke rehabilitation: the Willis lecture. Stroke, 31(1), pp.223-230.

[43] Veerbeek, J.M., van Wegen, E., van Peppen, R., van der Wees, P.J., Hendriks, E., Rietberg, M. and Kwakkel, G., 2014. What is the evidence for physical therapy poststroke? A systematic review and meta-analysis. PloS one, 9(2), p.e87987.

[44] Kong, K.H., Loh, Y.J., Thia, E., Chai, A., Ng, C.Y., Soh, Y.M., Toh, S. and Tjan, S.Y., 2016. Efficacy of a virtual reality commercial gaming device in upper limb recovery after stroke: a randomized, controlled study. Topics in stroke rehabilitation, 23(5), pp.333-340.
114

[45] Levin, M.F., Snir, O., Liebermann, D.G., Weingarden, H. and Weiss, P.L., 2012. Virtual reality versus conventional treatment of reaching ability in chronic stroke: clinical feasibility study. Neurology and therapy, 1(1), pp.1-15.

[46] Lange, B., Koenig, S., Chang, C.Y., McConnell, E., Suma, E., Bolas, M. and Rizzo, A., 2012. Designing informed game-based rehabilitation tasks leveraging advances in virtual reality. Disability and rehabilitation, 34(22), pp.1863-1870.

[47] Laver, K., George, S., Thomas, S., Deutsch, J. and Crotty, M., 2015. Virtual reality for stroke rehabilitation: an abridged version of a Cochrane review.

[48] Lake, C., 1997. Effects of prosthetic training on upper-extremity prosthesis use. Assessment, 8(11), p.12.

[49] Atzori, M. and Müller, H., 2015. Control capabilities of myoelectric robotic prostheses by hand amputees: a scientific research and market overview. Frontiers in systems neuroscience, 9, p.162.

[50] Ison, M. and Artemiadis, P., 2015. Proportional myoelectric control of robots: muscle synergy development drives performance enhancement, retainment, and generalization. IEEE Transactions on Robotics, 31(2), pp.259-268.

[51] van der Riet, D., Stopforth, R., Bright, G. and Diegel, O., 2013, September. An overview and comparison of upper limb prosthetics. IEEE, 2013 Africon, pp. 1-8.

[52] Bensmaia, S.J. and Miller, L.E., 2014. Restoring sensorimotor function through intracortical interfaces: progress and looming challenges. Nature Reviews Neuroscience, 15(5), pp.313-325.
115

classification control with a myoelectric prosthesis. Journal of neuroengineering and rehabilitation, 18(1), pp.1-15.